\documentclass[journal]{IEEEtran}

\usepackage[caption=false,font=normalsize,labelfont=sf,textfont=sf]{subfig}
\usepackage{mathrsfs}
\usepackage{latexsym}
\usepackage{graphicx}
\usepackage{epsfig}
\usepackage{subfig}
\usepackage{array}
\usepackage{amsmath}
\usepackage{amssymb}
\usepackage{color, soul}
\usepackage{amsthm}
\usepackage{enumerate}

\usepackage{algorithm}
\usepackage{algorithmic}

\usepackage{bm}
\usepackage{amssymb}
\usepackage{balance}
\usepackage{verbatim}
\usepackage{epstopdf}
\usepackage{setspace}
\usepackage{multicol}
\usepackage{amsmath,lipsum}
\usepackage{cuted}
\usepackage{multirow}
\usepackage{makecell}
\usepackage{url}

\begin{document}

\title{Anti-Jamming Sensing with Distributed Reconfigurable Intelligent Metasurface Antennas}

\author{Zhaowei Wang, Yunsong Huang, Weicheng Liu, Hui-Ming Wang, \emph{Senior Member, IEEE}
	

\thanks{The authors are with the School of Information and Communications Engineering, Xi'an Jiaotong University, Xi'an 710049, China, and also with the Ministry of Education Key Laboratory for Intelligent Networks and Network Security, Xi'an Jiaotong University, Xi'an 710049, China (e-mail: wangzwssd@stu.xjtu.edu.cn; song1102@stu.xjtu.edu.cn; liuweicheng@stu.xjtu.edu.cn; xjbswhm@gmail.com).}}



\maketitle

\begin{abstract}
The utilization of radio frequency (RF) signals for wireless sensing has garnered increasing attention. 
However, the radio environment is unpredictable and often unfavorable, the sensing accuracy of traditional RF sensing methods is often affected by adverse propagation channels from the transmitter to the receiver, such as fading and noise.
In this paper, we propose employing distributed Reconfigurable Intelligent Metasurface Antennas (RIMSA) to detect the presence and location of objects where multiple RIMSA receivers (RIMSA Rxs) are deployed on different places. 
By programming their beamforming patterns, RIMSA Rxs can enhance the quality of received signals.
The RF sensing problem is modeled as a joint optimization problem of beamforming pattern and mapping of received signals to sensing outcomes.
To address this challenge, we introduce a deep reinforcement learning (DRL) algorithm aimed at calculating the optimal beamforming patterns and  a neural network aimed at converting received signals into sensing outcomes. 
In addition, the malicious attacker may potentially launch jamming attack to disrupt sensing process.
To enable effective sensing in interference-prone environment, we devise a combined loss function that takes into account the Signal to Interference plus Noise Ratio (SINR) of the received signals.
The simulation results show that the proposed distributed RIMSA system can achieve more efficient sensing performance and better overcome environmental influences than centralized implementation.
Furthermore, the introduced method ensures high-accuracy sensing performance even under jamming attack.

\end{abstract}

\begin{IEEEkeywords}
RF sensing, Reconfigurable Intelligent Matasurface Antennas
(RIMSA), Deep reinforcement learning (DRL), beamforming, Anti-jamming.
\end{IEEEkeywords}

\section{Introduction}

\IEEEPARstart{R}{adio} Frequency sensing is a technology that utilizes radio frequency signals to detect and analyze the environment, objects, or human activities \cite{sense1}.
In the context of 6G wireless communication technology, RF sensing is explicitly regarded as one of the key technologies, with potential for achieving deep integration between communication and sensing. The vision for 6G aims to construct a comprehensive “intelligent connectivity of everything” network that spans across various dimensions. This network seeks to perceive the physical world and provide real-time feedback to the digital world.
The foundational principle of RF sensing is based on the analysis of changes in the propagation of wireless signals, such as variations in signal strength, phase, and doppler shift, to deduce information regarding the state of the surrounding environment or target objects.
RF sensing breaks through the physical limitations of traditional sensing technology through non-contact, penetrability, and environmental robustness, and is particularly suitable for privacy protection.
This technology has found increasingly broad application across various domains, such as security surveillance, health monitoring, intelligent transportation systems, and motion recognition \cite{sense2,sense3,sense4}. 

The wireless environment often poses challenges for RF sensing.
The inherent fading and noise can be detrimental to traditional RF sensing methods.
In recent years, Reconfigurable Intelligent Surfaces (RIS) has received increasing research as a solution to improve the environment and has a wide range of application scenarios, such as wireless communication \cite{risloc1, risloc2, risloc3, risloc4} and RF sensing \cite{senriszs1,senriszs2}.
These surfaces are able to apply varying phase shifts to RF signals, effectively mitigating the adverse effects of the wireless environment. Comprised of densely arranged electromagnetic (EM) excitation units, RIS can dynamically interact with EM fields.
This enables highly efficient shaping of EM waves to any intended purpose, with the ability to control, program, and customize wireless channels, thereby enhancing sensing accuracy \cite{ris1,ris2}.
RIS can significantly enhance the performance of wireless sensing and localization by actively modifying and controlling the electromagnetic properties of the surrounding environment \cite{senris3}.
In \cite{senris4}, the authors proposed a RIS based sensing system, where the RIS controller is employed to transmit pilot signals and sensing based on the received signals.
The authors in \cite{senris5} introduced an innovative RIS-assisted human activity recognition system that utilizes Wi-Fi signals.
In \cite{Hu1,Hu2}, the authors proposed a policy gradient based DRL algorithm to find the optimal beamforming configuration for sensing.
This approach not only reduces computational complexity but also enables adaptive and dynamic adjustment of beamforming pattern in response to changing environmental conditions.

In the above application scenarios,  RIS is primarily regarded as passive reflecting devices that assist traditional transceivers in shaping the propagation environment. 
Besides, RIS can also be used as antennas, which is referred to as Waveguide RIS or Dynamic Metasurface Antenna (DMA). 
The overall radiation mode of DMA is the superposition of radiation from all stimulated elements. The electromagnetic response of each metamaterial component can be altered to control the amplitude and phase of the radiation signal. The operation of each component can be programmed using a simple external electronic controller. 
Recently, DMA has also been applied in RF sensing and achieved favorable sensing performance \cite{dma1,dma2}.
In \cite{dma1}, the authors proposed a DMA based sensing framework and explained the antenna pattern diversity of the DMA provided a high-dimensional channel measurement to improve the sensing performance.
A RF sensing system was presented in \cite{dma2} that incorporated the DMA and the auxiliary-assisted ensemble multimask learning framework  to ensure high sensing accuracy.
Moreover, reconfigurable holographic surfaces (RHS) is an innovative type of planar antenna with densely deployed metamaterial elements, which is applied to the integrated sensing and communication \cite{rhs1,rhs2}. In \cite{rhs2}, the authors proposed a holographic beamforming scheme that jointly performed sensing
and communication.


Despite the discussions on using DMA and RHS to enhance wireless sensing performance, the aforementioned methods do not directly address how to improve the robustness of sensing in the presence of malicious jamming attack. 
The inherent broadcast nature and open characteristics of wireless channels render wireless transmissions vulnerable to jamming attacks \cite{secure1,secure2}. In such attack, malicious entities deliberately transmit jamming signals on legitimate channels, significantly degrading the performance of RF sensing system \cite{antijamming1,antijamming2,antijamming3}.
However, there is no research considering how to improve the accuracy of RF sensing, as the sensing accuracy is severely reduced due to the jamming attack.

In this paper, we propose an active reconfigurable intelligent metasurface antennas, referred to as RIMSA.
Compared with the serial activation of metamaterial elements in DMA and RHS, RIMSA is composed of metamaterial elements and a parallel feeding network, enabling the simultaneous and parallel activation of all elements through a coaxial feeding network. 
The reconfiguration of each metamaterial element is achieved by applying a DC voltage to varactor diodes, thus realizing continuous phase modulation. Multiple metasurface elements connect the RF chains to the digital signal processor, which controls the phase of each metasurface element. 
By integrating the functions of programmable RIS units and efficient power heat dissipation in a single module, a wide range of functionalities can be realized.
The encoding of the reconfigurable elements is referred to as the beamforming pattern of the metasurface. 
Furthermore,  to leverage the advantages of RIMSA, a hybrid beamforming approach is adopted, where analog beamforming is used to adjust the phase of the RIMSA, 
and a digital combiner is employed to further refine the amplitude and phase of the received signals \cite{dc1,dc2,dc3}.
Through dynamic design of the beamforming pattern and digital combiner, RIMSA can actively control RF signal beams during the sensing process.

We consider a RIMSA-assisted RF sensing scenario capable of detecting the presence and location of objects within a target space. 
To overcome unfavorable channel environments, RIMSA Rxs are placed on multiple walls, with each  receiver acting as one RF signal chain. 
Specifically, by programming the beamforming pattern of the receivers, it is possible to enhance the received signals, thereby providing high-quality sensing signals.
Achieving high sensing accuracy in a RIMSA-assisted RF sensing scenario faces the following challenges:
Firstly, it is essential to carefully design the beamforming pattern of the RIMSA Rxs to augment received signals that are advantageous for sensing.
Appropriate digital combining technique should be applied to merge these received signals.
Secondly, the mapping of the received signals into the detection of object presence and location, requires optimization to ensure accurate and reliable sensing outcomes.
Thirdly, the system must maintain high-precision sensing capabilities even in the presence of jamming attack.
To address these challenges, we introduce a DRL algorithm based on policy gradient method (defined as \textit{policy network}) for beam pattern selection \cite{antijammingai1,antijammingai2,antijammingai3}. 
Finding the optimal beamforming pattern is often formulated as an optimization problem. 
However, the complexity of identifying the optimal beamforming pattern is significantly high due to the associated optimization problem being a discrete nonlinear programming issue with a large number of variables.
To address this complexity, an increasingly popular approach is to model the control of these reconfigurable elements as an MDP and apply DRL for real-time coarse phase control.
Given fixed beam pattern,  we use the MRC method as a digital combiner to merge the received signals.
Then, the received signals are mapped to sensing outcomes through a neural network (defined as \textit{sensing network}).
The policy network and the sensing network are progressively optimized by minimizing the cross-entropy loss function of the sensing outcomes, achieving high-precision sensing.
When there is jamming attack during RF sensing, we propose an anti-jamming sensing strategy.
By optimizing the configuration of the RIMSA Rxs, the proposed method aims to cause the direct and reflected paths of the jamming signals to cancel each other out, thereby reducing the impact of jamming attack. Furthermore, we have redesigned the loss function to account for the SINR of the received signals, enabling the training of a robust sensing model upon completion. This approach enhances the system's robustness against jamming attack, maintaining high sensing accuracy in adversarial environment.



We summarize the contributions of this work as follows.
\begin{itemize}
	\item{
	To mitigate the impact of unfavorable wireless environments on RF sensing, we propose the utilization of distributed RIMSA Rxs to enhance sensing accuracy.
	The distributed RIMSA Rxs effectively extend the signal reception range and enhance RF sensing capabilities through spatial diversity.
	By adaptively configuring the phase states of RIMSA elements, the received signals can be coherently combined to mitigate signal fading and enhance the quality of the received signals.}
	\item{The programmable characteristics of RIMSA enable its integration with machine learning techniques, where neural networks can gradually acquire adaptive phase-shifting capabilities. At the RIMSA Rxs, we perform analog beamforming by utilizing the policy network to select the configurations. 
	The received signals are digitally combined to enhance the quality and then mapping to sensing outcomes through sensing network.
    Once the RF sensing process is initiated, the receiver learns the network model by processing received signals against ground-truth labels. After completing the training, the receiver can directly output sensing results. This framework achieves end-to-end RF sensing.}
	\item{In the face of potential jamming attack, we propose to redesign the loss function by integrating the SINR of the received signals with the cross-entropy loss function. 
	In this scenario, our objective is to concurrently accomplish RF sensing and jamming mitigation. Consequently, the loss function comprises two components: the cross-entropy term for achieving high-precision sensing and the SINR term that serves as an intuitive indicator of jamming rejection capability. This formulation demonstrates that the beamforming control ability of the policy network can position jamming sources within beam nulls, thereby enabling RIMSA to achieve spatial domain jamming suppression.
	Simulations confirm the effectiveness of our proposed method, demonstrating that sensing can be successfully achieved even under jamming attack.}

\end{itemize}

\textit{Notations}: $z$, $\bm{z}$ and $\bm{Z}$ stand for a
scaling factor, a column vector and a matrix, respectively.
$(\cdot)^H$, and $(\cdot)^T$ respectively denote conjugate transpose and transpose, $\mathbb{E}[\cdot]$ denotes the expected value of random
variable. $|\cdot|$ and $||\cdot||$denote the absolute value of a scalar and the Euclidean norm of a vector or matrix, respectively.  $\mathfrak{R}$ denotes the real part of a complex number.
$\bm{0}_k$, and $\bm{1}_k$ denote all-zero and all-one  column vectors with $k$-dimensions, respectively.


\begin{figure}[!t]
	\centering
	\includegraphics[width=3.2in]{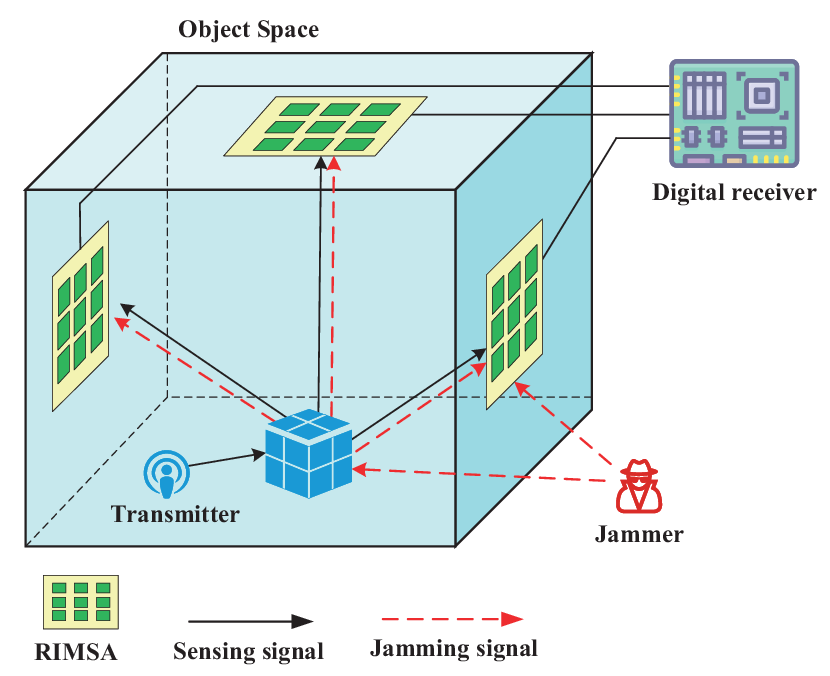}
	\caption{System model.}
	\label{system model}
\end{figure}
\section{System Model}
We assume a target space that contains the object of interest. In this scenario, a single-antenna transmitter (Tx) is employed to emit probe signals for sensing the target object. Each RIMSA is placed in different direction of the target space. The target space is modeled as a cubic region, discretized into $M$ equally sized spatial grid points, where each grid point has dimensions of $\triangle l_x\times \triangle l_y\times \triangle l_z$. The transmitter emits a signal at a carrier frequency of $f_c$. 
We assume the digital receiver is capable of adjusting the amplitude and phase of each RIMSA. 
The digital receiver and each RIMSA are connected via wired links. This allows for both analog and digital beamforming. 
Each RIMSA adjusts its phase configuration to control the reception of the signal. Then, all received signals are processed through a digital combiner to generate the final received signal. This signal is then used to enable the sensing of the target object.
We assume that each RIMSA has $N$ reconfigurable elements, and each reconfigurable element has $N_s$ phase configurations. Every RIMSA forms a single RF chain, with a total of $N_{RF}$ RF chains.

Furthermore, we assume the presence of an attacker attempting to launch a jamming attack to disrupt the sensing process. The attacker is located within a specific region, but the exact position is not fixed.
This digital receiver enables coordinated and adaptive beamforming, which is crucial for effective signal processing and jamming mitigation in complex wireless environment.
The digital receiver can dynamically configure the parameters of the RIMSA to optimize the beamforming pattern, thereby enhancing the overall performance of the sensing system. 

\begin{figure}[!t]
	\centering
	\includegraphics[width=3.2in]{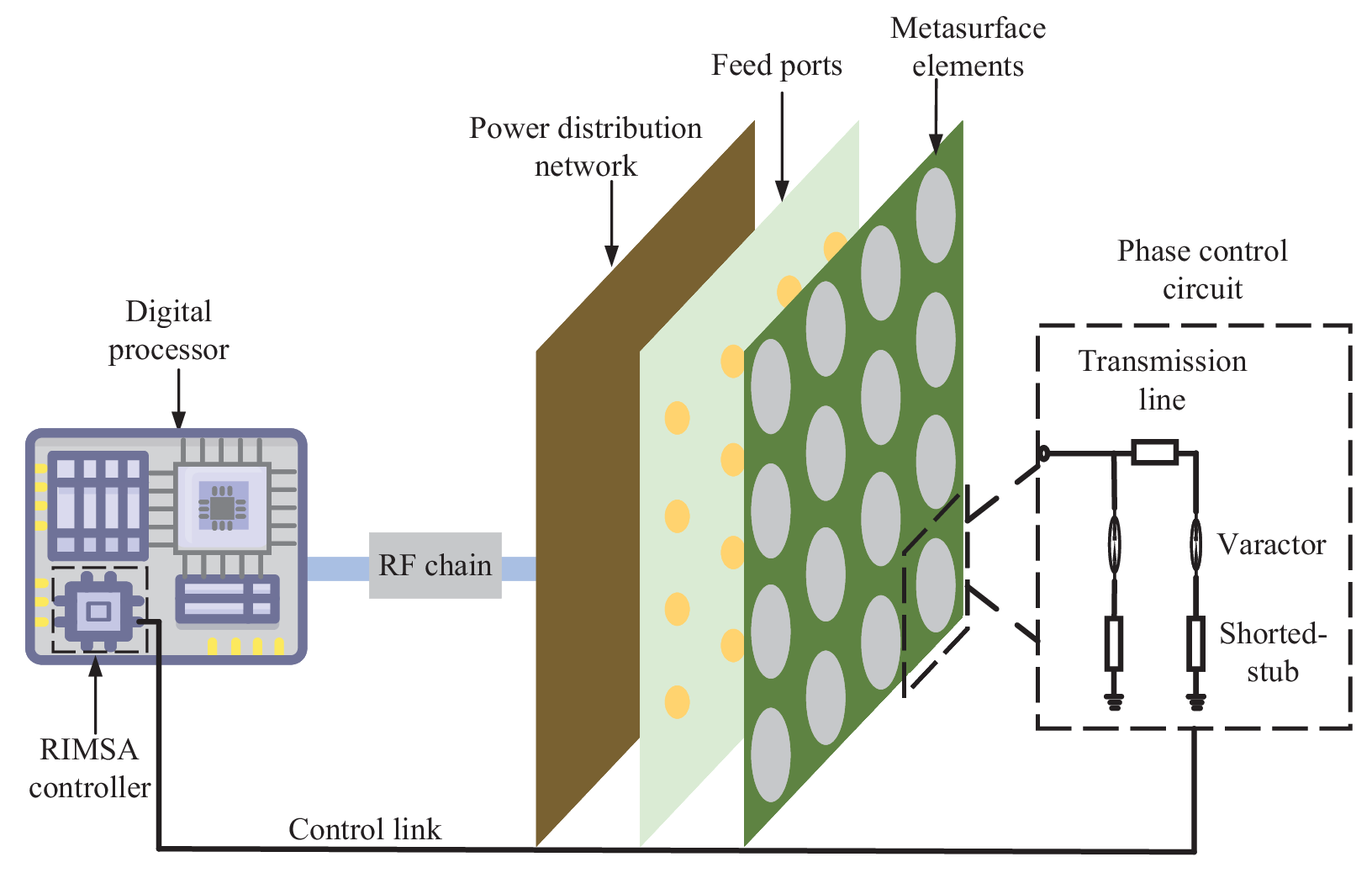}
	\caption{The architecture of a RIMSA.}
	\label{RIMSA}
\end{figure}

\subsection{RIMSA Structure}
The RIMSA is a flexibly configurable metasurface antenna,
allowing for adaptation and reconfiguration according to the
wireless environment. The architecture of RIMSA is shown in Fig. \ref{RIMSA}.
One RIMSA is consists of phase control circuits, metamaterial elements, power distribution network and feed ports. The phase control circuit
of each metamaterial element is responsible for adjusting
the phase of EM signal, which is composed of a varactor,
transmission line and short stub. Specifically, by regulating
the DC voltage of the phase control circuis, the parameter
of varactors can be adjusted to change the phase response
of the metamaterial element. The phase shifting is contingent
upon phase control circuit, which leads to a continuous phase control of the EM wave and analog beamforming. The RF
power is distributed to each metamaterial element uniformly
by power distribution network, which is implemented by
means of microstrip line power dividers. Feed ports contains
a main feed port as well as ports for power divider. These
four principal parts are tightly connected, which ensures a
compact antenna architecture. All the metamaterial elements
are excited simultaneously by the impinging signal, and arrive
at the RF chain and sequentially the digital processor through
a parallel feed network, which avoids the frequeny selectivity
of propagation in microstrip line as in a DMA.

This structure enables dynamic integration of multipath signals, demonstrating strong performance in two typical scenarios: anti-fading and anti-jamming.
In the anti-fading scenario, the RIMSA performs coherent signal combining from multiple propagation paths using its large number of tunable elements.
In the anti-jamming scenario, the RIMSA forms nulls in specific directions by adjusting the phase of each element to suppress signals coming from interferers.
By appropriately designing the phase shift of each unit, the RIMSA is capable of signal enhancement and interference cancellation in complex electromagnetic environments.

\begin{figure}[!t]
	\centering
	\includegraphics[width=3.2in]{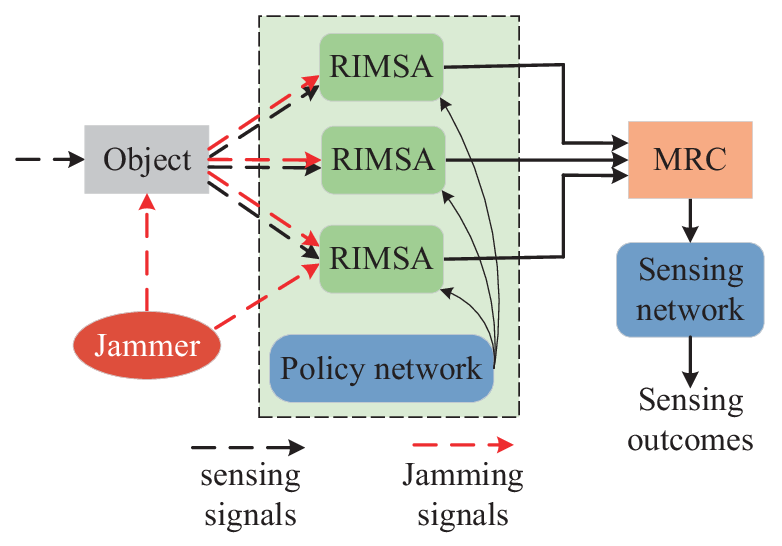}
	\caption{The signal processing flow in RF sensing.}
	\label{overview}
\end{figure}

\subsection{Sensing Procedure}

The complete processing flow for RF sensing signals is illustrated in Fig. \ref{overview}. The transmitter continuously emits directional signals towards objects, which are then reflected to RIMSA units located at different positions. In the presence of an attacker, jamming signals emitted by the attacker can also be reflected from the object to the RIMSA units, thereby interfering with the normal sensing signals. 
In this work, we adopt line-of-sight (LoS) path channel model to evaluate the performance of the proposed distributed RF sensing system and verify the feasibility of the anti-jamming approach \cite{los1,los2}. 
To validate the proposed anti-jamming method, we assume a powerful attacker that also employs a directional antenna to target the object while simultaneously being capable of directly attacking the nearest RIMSA. The jamming signal follows a LoS path, exhibiting strong interference capability.
Within the digital receiver, a policy network dynamically adjusts the phase configuration of the RIMSA to achieve higher quality received signals. This policy network employs reinforcement learning based on policy gradient algorithms. Subsequently, the digital receiver combines the signals received by the RIMSA and feeds them into a sensing network. This sensing network is implemented using a multi-layer perceptron (MLP) to produce the final sensing outcomes.

\begin{figure}[!t]
	\centering
	\includegraphics[width=3.2in]{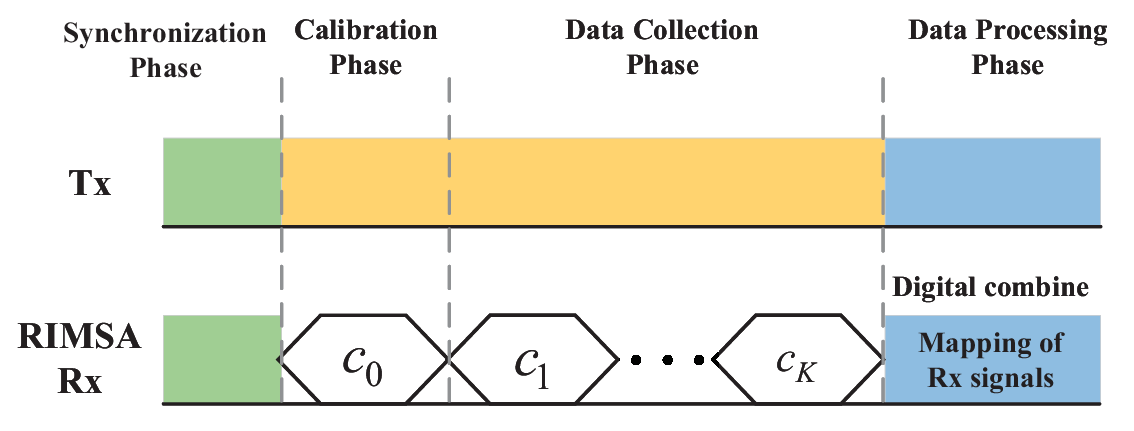}
	\caption{RF sensing protocol.}
	\label{protocol}
\end{figure}

To clearly describe the process of RIMSA-assisted RF sensing, the following RF sensing procedure is established. In this protocol, time is divided into time slots and organized into periods, where the Tx and RIMSA operate in a synchronized and periodic manner. 
As shown in Fig. \ref{protocol}, each period is comprised of four phases: synchronization, calibration, data acquisition, and data processing.
During the synchronization phase, the Tx emits a synchronization signal to initiate the cycle. Following this, in the calibration stage, the Tx sends out a narrowband signal, represented as $x$, at the carrier frequency $f_c$. 
Moreover, we refer to the vector of configurations selected
for the $N$ reconfigurable elements within one RIMSA as a beamformer pattern, which can be represented by an $N\cdot N_s$-dimensional binary vector ${\bm{c}}=(\bm{\hat{o}}(c_1)^T,\ldots,\bm{\hat{o}}(c_N)^T)^T\in \mathbb{C}^{N\cdot N_s \times 1 }$, where
$\bm{\hat{o}}(c_n)^T\in \mathbb{C}^{N_s \times 1 }, \forall n\in [1,N]$ denotes the $N_s$-dimensional vector whose $c_n$-th element is $1$ and the other elements are all $0$. Specifically, $c_n \in [1,N_s]$ denotes the configuration of the $n$-th reconfigurable element and $c_n \in \mathbb{Z}$ where $\mathbb{Z}$ denotes the set of integers.
The beamforming mode of all RIMSA is represented as $\bm{C}=(\bm{c}^1,\ldots,\bm{c}^{n_{RF}},\ldots,\bm{c}^{N_{RF}})^T\in \mathbb{C}^{ N_{RF}\times N\cdot N_s }$.
Here, we refer to $\bm{C}$ as the control matrix of RIMSA and $n_{RF}\in[1,N_{RF}]$ denotes $n_{RF}$-th RF signal chain.

The above descriptions explain the control matrix of distributed RIMSA in one frame signal during the data collection phase.
The data acquisition phase is segmented into $K$ frames equally.
The $K$ beamforming modes of all RF chains during the data acquisition phase form the control matrix $\vec{\bm{C}}$, which can be expressed as $\vec{\bm{C}}=(\bm{C}_1,\bm{C}_2,\ldots,\bm{C}_K)$.
The policy network from the reinforcement learning module selects the new configuration for each of the reconfigurable elements at the end of each frame.  
Throughout this phase, the Tx maintains the transmission of the narrowband RF sensing signal, and at the end of each frame, the RIMSA adjusts its beamforming mode to a new configuration.
The signal received at the digital receiver is denoted as $\bm{y}$.  

In the following data processing stage, the received signal needs to undergo further amplitude and phase control at the digital receiver. 
We use MRC for merging all RF chains to improve the signal-to-noise ratio (SNR) of the received signal. Then, the merged signal will be fed into sensing network to obtain the sensing outcomes.

\subsection{Received Signals}
\subsubsection{Received Signals Without Jamming Attack}

In the considered RIMSA-assisted RF sensing scenario, the Tx is equipped with a single antenna to transmit RF signals. The transmitting antenna is a directional antenna aimed at the target object. As illustrated in the Fig. \ref{system model}, the Tx emits a RF signal, which is reflected by the target object and then reaches the RIMSA Rx. 
For each RF link, the channel can be described as

\begin{equation}
	\begin{split}
		h_{m,n}^{Tx,Obj,Rx}\left( v_m,c_n \right) &= \frac{{\lambda} ^2 \cdot {v_m} \cdot r_{m,n}\left( {c_n} \right) \cdot \sqrt {{g_T}} }{{{\left( {4\pi } \right)}^2} \cdot {d_{m}^{Tx,Obj}} \cdot {d_{m,n}^{Obj,Rx}}} \\
		& \cdot {e^{ - j2\pi \left( {{d_{m}^{Tx,Obj}} + {d_{m,n}^{Obj,Rx}}} \right)/\lambda }}, 
	\end{split}
\end{equation}
$\lambda$ is the wavelength of the signal, $g_{T}$  represents the gain of the Tx. $d_{m}^{Tx,Obj}$ is the distance from Tx to the $m$-th grid point of the object. $d_{m,n}^{Obj,Rx}$ is the distance from the $m$-th grid point of the object to $n$-th element of RIMSA Rx.
Use $r_{m,n}(c_n)$ to represent the reflection coefficient of the $n$-th reconfigurable element.
The reflection coefficient of the $m$-th spatial grid is $v_m$.
For each RF chain, the received signal can be expressed as
\begin{equation}\label{RFsingley}
	y = \sum\limits_{n=1}^{N} {\sum\limits_{m = 1}^M {{h_{m,n}^{Tx,Obj,Rx}}\left( {{v_m},{c_n}} \right) \cdot \sqrt {{P_{Tx}}}  \cdot x + } } \sigma, 
\end{equation}
$\sigma$ denotes the additive white Gaussian noise with power $\epsilon$.
The vector of configurations chosen for the $N$ reconfigurable elements is referred to as a beamformer pattern of the metasurface.
Based on the  this definition, the received signal in \eqref{RFsingley} can be rewritten as


\begin{equation}
	y= \bm{c}^T \cdot \bm{A} \cdot \bm{v} \cdot \sqrt{P_{Tx}}\cdot x+ \sigma,
\end{equation}
where $\bm{v}=(v_1,\ldots,v_M)^T\in \mathbb{C}^{ M\times 1}$ denotes the vector of reflection coefficients of the $M$ space grids.
$\bm{A} =(\bm{\alpha}_1,\ldots,\bm{\alpha}_M) \in \mathbb{C}^{N\cdot N_s \times M }$ is referred to as the projection matrix, and $\bm{\alpha}_m =(\bm{\hat{\alpha}}^T_{m,1},\ldots,\bm{\hat{\alpha}}^T_{m,N} )^{T}\in \mathbb{C}^{N\cdot N_s \times 1 }$ with $\bm{\hat{\alpha}}_{m,n} =(\hat{\alpha}_{m,n,1},\ldots,\hat{\alpha}_{m,n,N_s} )^T \in \mathbb{C}^{ N_s \times 1}$. 
Specifically, for all $m \in [1,M]$, $n \in [1,N]$, and $i \in [1,N_s]$, $\hat{\alpha}_{m,n,i}$ represents the channel gain of the reflection path through the $n$-th reconfigurable element when it is in configuration $i$, corresponding to the $m$-th spatial grid point, assuming a unit reflection coefficient. This channel gain can be mathematically expressed by the equation
\begin{equation}
	\begin{split}	
		\hat{\alpha}_{m,n,i} =& \frac{{\lambda} ^2 \cdot r_{m,n}\left( {i} \right) \cdot \sqrt {{g_T}}\cdot {e^{ - j2\pi \left( {{d_{m}^{Tx,Obj}} + {d_{m,n}^{Obj,Rx}}} \right)/\lambda }} 
		}{{{\left( {4\pi } \right)}^2} \cdot {d_{m}^{Tx,Obj}} \cdot {d_{m,n}^{Obj,Rx}}}. 
	\end{split}
\end{equation}

We assume there are $N_{RF}$ RIMSA Rxs, which means there are $N_{RF}$ RF chains.
Each control matrix independently regulates the received signal of a RIMSA Rx, which can be represented as 
$\bm{C}= \left(\bm{c}^1,\ldots,\bm{c}^{N_{RF}}\right)\in \mathbb{C}^{N_{RF}\times  N \cdot N_s}$.
The received signal can be represented as
\begin{equation}\label{bigA}
	\bm{y}= \bm{C} \cdot \bm{A} \cdot \bm{v} \cdot \sqrt{P_{Tx}}\cdot x+ \bm{\sigma}.
\end{equation}

To improve the quality of the received signal, especially in the presence of multipath signals, we use MRC to linearly combine the signals received over multiple RF chains to augment the quality of the signals at the receiver, thereby improving the overall performance of the system. 
We define the equivalent channel after RIMSA phase manipulation as
\begin{equation}
	\bm{h}= \bm{C} \cdot \bm{A} \cdot \bm{v}.
\end{equation}
The digital combiner at the receiver that employs MRC can be defined as follows
\begin{equation}
	\bm{w}^T_{MRC}=\frac{\bm{h}^H}{||\bm{h}||^{2}}.
\end{equation}
The combined received signal can be represented as
\begin{equation}\label{mear mtrix}
	\hat{y}=\bm{w}^T_{MRC}\cdot \bm{y} = \bm{w}^T_{MRC}\cdot \bm{C} \cdot \bm{A} \cdot \bm{v} \cdot \sqrt{P_{Tx}} \cdot x+\hat{\sigma},
\end{equation}
where $\hat{\sigma}=\bm{w}^T_{MRC}\cdot \bm{\sigma}$ is the noise after digital combining.
According to the sensing protocol, we will send $K$ frames of signals for RIMSA phase adjustment within one period.
we denote the beamformer patterns of the RIMSA corresponding to the $K$ frames by the binary row vectors $\bm{C}_1,\ldots, \bm{C}_K$. Specifically, the $K$ beamformer patterns of the RIMSA during the data collection phase constitutes the control matrix, which is denoted by $\bm{\vec{C}}=(\bm{C}_1,\ldots, \bm{C}_K)$. 
In order to clearly see the relationship between the received signal and the reflection coefficient of the object from \eqref{mear mtrix}, we define $(\bm{\gamma}^{Tx}_k)^T = \bm{w}^T_{MRC}\cdot \bm{C}_k\cdot \bm{A}\cdot \sqrt{P_{Tx}} \cdot x\in \mathbb{C}^{1\times M}$.
In this way, the combined signals of all RF chains in $k$-th frame can be simplified as $\hat{y}_k=(\bm{\gamma}^{Tx}_k)^T\cdot \bm{v} +\hat{\sigma}$.
Hence, the received signals will form a vector $\bm{\hat{y}}\in \mathbb{C}^{K\times 1}$ as
\begin{equation}
\bm{\hat{y}}=\bm{\Gamma}_{Tx} \cdot \bm{v} +\hat{\bm{\sigma}},
\end{equation}
where $\bm{\Gamma}_{Tx} = (\bm{\gamma}^{Tx}_1,\ldots,\bm{\gamma}^{Tx}_K)^T  \in \mathbb{C}^{K\times M}$.

\subsubsection{Received Signals under    Jamming Attack}
Considering the presence of an attacker in the scenario, and given that the position of the attacker is difficult to track, we assume that it moves randomly within a cubic region. In this case, using the aforementioned sensing model would significantly reduce the sensing accuracy due to the jamming attack.

Due to the presence of the attacker, the reflected path of the jamming signal need to be accounted for 
\begin{equation}
	\begin{split}
		h_{m,n}^{J,Obj,Rx} (c_n,v_m)  =\ & 
		\frac{\lambda ^2 \cdot r_{n,m}(c_n)\cdot v_m \cdot \sqrt{g_J}}{(4\pi)^2 \cdot d_{m}^{J,Obj}\cdot d_{m,n}^{Obj,Rx}} \\
		& \cdot {e^{ - j2\pi \left( d_{m}^{J,Obj} + d_{m,n}^{Obj,Rx} \right)/\lambda }}, \\
	\end{split}
\end{equation}
$d_{m}^{J,Obj}$ is the distance from the attacker to the $m$-th grid point of the object. $d_{m,n}^{Obj,Rx}$ is the distance from the $m$-th grid point of the object to $n$-th element of RIMSA Rx.
Furthermore, in addition to the reflection paths targeting the object, we consider that an attacker may attempt to jam the RF link. The distributed architecture of the system makes such an attack more feasible, as illustrated in Fig. \ref{system model}.
The direct path of the jamming signal need to be accounted for
\begin{equation}
	h_{n}^{J,los} =\frac{\lambda}{4\pi}\cdot\frac{\sqrt{g_J}\cdot e^{-j2\pi d_n^{J,los}/\lambda}}{d_n^{J,los}},
\end{equation}
where $d_n^{J,los}$ denotes the distance from the attacker to the RIMSA Rx.
We assume that the attacker only launches jamming attack on one piece of RIMSA, so the LoS path attack can be written as $\bm{h}_{J,los}\in \mathbb{C}^{N_{RF}\times 1}$. It is a $N_{RF}$-dimensional all zero column vector, where the $n_{RF}$-th element is the specified by $\sum_{n=1}^{N}h_{n}^{J,los}$.
Hence, the received signals of all RF links under jamming attack can be formulated as
\begin{equation}
	\begin{split}
		\bm{y}_J=\ & \bm{C} \cdot \bm{A} \cdot \bm{v} \cdot \sqrt{P_{Tx}}\cdot x +\bm{C} \cdot \bm{A}_J \cdot \bm{v} \cdot \sqrt{P_{J}}\cdot z + \\
		&\bm{h}_{J,los} \cdot \sqrt{P_{J}}\cdot z + \bm{\sigma},
	\end{split}
\end{equation}
where $\bm{A}_J =(\bm{\alpha}_1,\ldots,\bm{\alpha}_M ) \in \mathbb{C}^{ N\cdot N_s \times M }$ is referred to as the projection matrix of the attacker, which is similar to the definition in \eqref{bigA}.
Then, the combined received signal under jamming attack can be represented as
\begin{equation}
	\begin{split}
		\hat{y}_J 
		=\ &\bm{w}^T_{MRC}\cdot \bm{C} \cdot \bm{A} \cdot \bm{v} \cdot \sqrt{P_{Tx}}\cdot x + \\
		&\bm{w}^T_{MRC}\cdot\bm{C} \cdot \bm{A}_J \cdot \bm{v} \cdot \sqrt{P_{J}}\cdot z + \\
		&\bm{w}^T_{MRC}\cdot \bm{h}_{J,los} \cdot \sqrt{P_{J}}\cdot z + \hat{\sigma},
	\end{split}
\end{equation}

Similarly, we define $(\bm{\gamma}^{J}_k)^T = \bm{w}^T_{MRC}\cdot \bm{C}_k\cdot \bm{A}_J \cdot \sqrt{P_{J}}\cdot z\in \mathbb{C}^{1\times M}$.
In this way, the combined received signals of all RF chains in $k$-th frame can be simplified as $\hat{y}_J=(\bm{\gamma}^{Tx}_k)^T\cdot \bm{v}+(\bm{\gamma}^{J}_k)^T\cdot \bm{v} + \bm{w}^T_{MRC}\cdot \bm{h}_{J,los} \cdot \sqrt{P_{J}}\cdot z + \hat{\sigma}$.
Hence, the received signals will form a vector
\begin{equation}
	\bm{\hat{y}}_J=\bm{\Gamma}_{Tx} \cdot \bm{v} + \bm{\Gamma}_{J} \cdot \bm{v}+ \bm{y}_{J,los} + \hat{\bm{\sigma}},
\end{equation}
where $\bm{\Gamma}_{J} = (\bm{\gamma}^{J}_1,\ldots,\bm{\gamma}^{J}_K )^T \in \mathbb{C}^{K\times M}$ and $\bm{y}_{J,los}=\bm{w}_{MRC}\cdot \bm{h}_{J,los} \cdot \bm{1}_K \cdot \sqrt{P_{J}}\cdot z\in \mathbb{C}^{K\times 1} $.

\section{RIMSA Control and Sensing Optimization Problem}
In the data processing phase, the digital receiver transforms the measurement vector acquired during the data acquisition phase into sensing outcome. This result manifests as a probability vector that signifies the likelihood of an object being present at each of the 
$M$ spatial grid points.
This mapping relationship can be expressed as a parametric function, $\bm{\hat{p}}=\bm{f}^{\bm{w}}(\bm{y})$.
This function, referred to as the mapping of the received signal, is realized through a neural network called sensing network and parameterized by the vector $\bm{w}$. 

The sensing outcome is determined by $\bm{f}^{\bm{w}}(\bm{\hat{y}})$.
In this work, we consider $\bm{f}^{\bm{w}}(\bm{\hat{y}})$ to be a nonlinear function modeled as a neural network, where the elements of $\bm{w}$ represent the weights of the nodes within the network.
To optimize the metasurface-assisted scenario for maximum sensing accuracy, we formulate the problem as a following loss minimization task. In this formulation, the control matrix $\bm{\vec{C}}$ and the parameters of the mapping function for the received signals $\bm{w}$ serve as the optimization variables. The objective is to minimize the loss between the predicted outcomes from the sensing network and the true labels, thereby enhancing the accuracy of the sensing results.
In addition, the mapping result $\bm{\hat{p}}$ will be presented in the form of a probability vector with a dimension of $M$.
Each element  $p_m \in[0,1]$, represents the probability of an object existing at the $m$-th grid point.
Consequently, $1-p_{m}$ signifies the probability that the  $m$-th spatial grid point is unoccupied.


In an environment without jamming attack, we use the cross-entropy loss function as the objective function to train the network. The use of cross entropy loss function can show the accuracy of the network's perception during training. This is because as the loss decreases, the network's predictions become closer to the real situation.
The cross-entropy loss defined in the considered scenario can be expressed as
\begin{equation}
	L_{CE}=-E_{\bm{v}\in \mathcal{V}}[\sum_{m=1}^{M}p_m(\bm{v})\cdot \ln(\hat{p}_m)+(1-p_m(\bm{v}))\cdot \ln(1-\hat{p}_m)],
\end{equation}
Here, $\mathcal{V}$ represents the set of all possible reflection coefficient vectors corresponding to the presence of objects in the target space, while $p_m(\bm{v})$ is a binary variable indicating whether an object exists in the 
$m$-th spatial grid point. Specifically, $p_m(\bm{v})$ can be expressed as
\begin{equation}   
	p_{m}(\bm{v}) =
	\begin{cases}
		0,   &  \text{if $|v_m|=0$,} \\
		1,   &  \text{otherwise.}
	\end{cases}                
\end{equation}

When there is no jamming attack, the sensing problem can be formulated as
\begin{align}\label{P1}
	& (P1):  \mathop{\min}_{\bm{\vec{C}},\bm{w}} \, L_{CE}(\bm{\vec{C}},\bm{w}), \\
	& \quad s.t.  \ \ (\hat{p}_1,\cdots,\hat{p}_M)=\bm{f}^{\bm{w}}(\bm{\hat{y}}), \\
	& \qquad \quad \bm{\hat{y}}=\bm{\Gamma}_{Tx}\cdot \bm{v}+\hat{\bm{\sigma}}, \\
	& \qquad \quad \bm{\Gamma}_{Tx} = (\bm{\gamma}^{Tx}_1,\ldots,\bm{\gamma}^{Tx}_K)^T ,  \\
	& \qquad \quad (\bm{\gamma}^{Tx}_k)^T = \bm{w}_{MRC}\cdot \bm{C}_k\cdot \bm{A}\cdot \sqrt{P_{Tx}} \cdot x, \\
	& \qquad \quad \bm{\vec{C}}=(\bm{C}_1,\ldots, \bm{C}_K). 
\end{align}

When jamming attack is present, considering only the cross-entropy loss function may fail to achieve high-accuracy sensing, and can even lead to scenarios where the loss function does not decrease, causing the algorithm to fail to converge. To address this issue, we propose integrating the SINR with the cross-entropy loss to construct a new loss function. The SINR serves as an indicator of the control network's capability to mitigate jamming attack at the RIMSA Rxs. As the SINR increases, the impact of jamming decreases, thereby improving the system's performance.
The SINR can be calculated as
\begin{equation}
	\text{SINR} = \frac{P_{Tx}||\bm{\Gamma}_{Tx}\cdot \bm{v}||^2}{P_{J}||\bm{\Gamma}_{J}\cdot \bm{v}+\bm{h}_{J,los}||^2+\hat{\bm{\sigma}}^2}.
\end{equation}
To mitigate the jamming attack, we define a combined loss function to be applied during the training phase. 
The goal is to achieve RF sensing and anti-jamming performance simultaneous. Accordingly, the loss function comprises two components: 1) a cross-entropy term for high-accuracy sensing, and 2) a SINR term that quantifies anti-jamming capability. 
The last term explicitly represents  beamforming control ability of the policy network, which can locate the jamming signal within the beam nulls.
By employing this combined loss function during training, the resulting network exhibits enhanced robustness against jamming attack.
The optimization problem can be formulated as
\begin{align}\label{P2}
	&(P2): \mathop{\min}_{\bm{\vec{C}},\bm{w}} \, L_{CE}(\bm{\vec{C}},\bm{w})-\beta \log(1+\text{SINR}), \\
	& \quad s.t.  \ \ (\hat{p}_1,\cdots,\hat{p}_M)=\bm{f}^{w}(\bm{\hat{y}}), \\
	& \qquad \quad \bm{\hat{y}}=\bm{\Gamma}_{Tx}\cdot \bm{v}+  \bm{\Gamma}_{J}\cdot \bm{v} + \bm{y}_{J,los} + \hat{\bm{\sigma}}, \\
	& \qquad \quad \bm{\Gamma}_{Tx} = (\bm{\gamma}^{Tx}_1,\ldots,\bm{\gamma}^{Tx}_K )^T,  \\
	& \qquad \quad \bm{\Gamma}_{J} = (\bm{\gamma}^{J}_1,\ldots,\bm{\gamma}^{J}_K )^T,  \\
	& \qquad \quad (\bm{\gamma}^{Tx}_k)^T = \bm{w}_{MRC}\cdot \bm{C}_k\cdot \bm{A}\cdot \sqrt{P_{Tx}} \cdot x, \\
	& \qquad \quad (\bm{\gamma}^{J}_k)^T = \bm{w}_{MRC}\cdot \bm{C}_k\cdot \bm{A}_J\cdot \sqrt{P_{J}} \cdot z, \\
	&\qquad \quad \bm{y}_{J,los}=\bm{w}_{MRC}\cdot \bm{h}_{J,los} \cdot \bm{1}_K \cdot \sqrt{P_{J}}\cdot z, \\
	& \qquad \quad \bm{\vec{C}}=(\bm{C}_1,\ldots, \bm{C}_K). 
\end{align}


In (P1) and (P2), our objective is to minimize \eqref{P1} and \eqref{P2} by optimizing 
$\bm{\vec{C}}$ and $\bm{w}$. 
After multiple epochs of training, we gradually optimized the parameter $\bm{\vec{C}}$ and $\bm{w}$ by minimizing the loss function \eqref{P1} and \eqref{P2}.
However, given $\bm{w}$, the problem remains challenging due to the large number of integer variables in the control matrix 
$\bm{\vec{C}}$. Additionally, a substantial number of training epochs may be required before converging to a local optimal solution. 
To efficiently solve (P1) and (P2), we model the problem as a Markov process  and employ DRL to address it.
This formulation leverages the sequential decision-making capabilities of DRL, which is well-suited for handling the dynamic and stochastic nature of the problem. By framing the problem within an MDP framework, we can define the state space, action space, transition probabilities, and reward function, allowing the DRL to learn an optimal policy through interaction with the environment.

\section{Solution to RF sensing optimization problem}
The overall process of RF sensing from the transmitted signal to the received signal consists of two main components, as shown in Fig. 3. 
The first component is responsible for guiding the RIMSA to select appropriate beamforming pattern (phase configuration of all metasurface elements) to enhance the received signal quality and anti-jamming. This part is realized through DRL, where the network used in this context is referred to as the policy network. The second component is tasked with mapping the received signals to the sensing outcomes, which is modeled using a neural network and is referred to as the sensing network.
Here, we adopt an end-to-end learning approach inspired by machine learning techniques, where the goal is to directly obtain sensing results while implicitly embedding parameter analysis in the training of network parameters. 
This approach encapsulates the analysis of reflection coefficients and received signals within the network parameters, enabling direct and efficient inference of the sensing outcomes.

\begin{figure}[!t]
	\centering
	\includegraphics[width=3.2in]{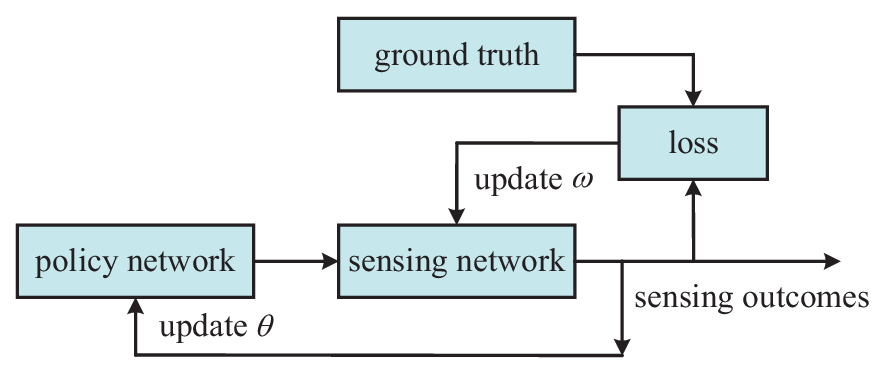}
	\caption{The training process of the policy network and sensing network.}
	\label{train procedure}
\end{figure}

During the training phase, the strategy network updates its parameters by calculating the gradient of the value function based on accumulated rewards. This involves using gradient descent methods to optimize the network, thereby improving its ability to select optimal beamforming pattern.
Then, the sensing network calculates the error between the predicted sensing outcomes and the true labels using a cross-entropy loss function. Back propagation (BP) algorithm is then used to compute gradients and update the network parameters, optimizing its performance for accurate sensing.
The training process is shown in the Fig. \ref{train procedure}.
This dual-component approach leverages the strengths of both reinforcement learning and neural networks to optimize the RF sensing process. The policy network, through reinforcement learning, dynamically adjusts the beamforming pattern to adapt to the environment and mitigate the jamming attack. Meanwhile, the sensing network processes the received signals to extract meaningful information, enabling accurate and reliable sensing.

\subsection{Configuration Selection Based on Policy Network}
We assume that the digital receiver has full control over all RIMSA Rx units and can determine the configuration of each reconfigurable element in every beamforming mode. (P1) and (P2) can thus be framed as a decision optimization problem, which can be effectively addressed using DRL algorithm. Since DRL requires the target problem to be modeled as a MDP, it involves four key components: the state set $\mathcal{S}$, the set of available actions $\mathcal{A}$, the state transition function $\mathcal{T}$, and the reward function $\mathcal{R}$.



State: In the distributed RIMSA-assisted RF sensing scenario, the state is defined in the form of 
\begin{equation}
	\bm{s}=(k,n,\bm{C}),
\end{equation}
where $k$ and $n$ are the row and column indices of the control matrix $\bm{C}$, representing the configuration that the RIMSA intends to select. 
The initial state of is denoted as $\bm{s}_0=(1,1,\bm{C}_0)$
, where $\bm{C}_0$ represents the initial control matrix for the RIMSA, with all reconfigurable elements set to their first configuration. This starting point signifies the beginning of the decision-making process.
The state indexed by $(k,n)=(K+1,1)$ is referred to as the terminal state. 
Upon reaching this terminal state, it indicates that all configurations in the control matrix have been sequentially selected and applied.

Action: In each state $\bm{s}$, the RIMSA selects a configuration for the $n$-th reconfigurable element in the $k$-th frame.
The action set available to the RIMSA in each state can be represented as $\mathcal{A}=\{1,\ldots,N_s\}$, where the $j$-th  $(j\in[1,N_s])$
indicates that the RIMSA chooses to set the target element to its $j$-th configuration. Thus, at each step within a given state, the digital receiver has the flexibility to choose from $N_s$ distinct actions, enabling it to explore different configurations for improving the sensing performance.

State Transition Function: After the digital receiver selects the action, the MDP framework transits into the next state $\bm{s'}=(k',n',\bm{C}_{next})$.
If $(k, n)$ = $(K,N)$, the receiver enters the terminal state of the MDP. For the non-terminal states, the elements of state $\bm{s'}$ given
$\bm{s}$ and $a$ can be expressed as follows
\begin{equation}
	k'=k+1,n'=\text{mod}(n,N)+1.
\end{equation}
\begin{equation}
    (\bm{C}_{next})_{(k'',n'')}=
    \begin{cases}
    	(\bm{C}_{next})_{(k'',n'')}, \: \text{if}\: (k'',n'')\neq(k,n), \\
    	\hat{\bm{o}}(a), \: \text{if}\: (k'',n'')=(k,n).
    \end{cases}    
\end{equation}
The control matrix at the termination state is represented by $\bm{C}_t$.

Reward: At the terminal state, the reward is defined as the negative loss function. This means that upon reaching the terminal state, where all configurations have been selected, the reward is calculated based on how well the sensing performance has been optimized.
If the terminal state has not yet been reached, the reward for each state transition is set to zero. This approach focuses the learning process on optimizing the final outcome rather than rewarding intermediate steps, ensuring that the chosen actions leading to optimal configurations at the end of the sequence.
The reward in state $\bm{s}$ is defined as
\begin{equation}   
	R(\bm{s}|\bm{w}) =
	\begin{cases}
		-L_{CE}(\bm{C}_t,\bm{w}),   &  \text{if $\bm{s}$ is a terminal state,} \\
		0,   &  \text{otherwise.}
	\end{cases}                
\end{equation}

\begin{figure}[!t]
	\centering
	\includegraphics[width=3.2in]{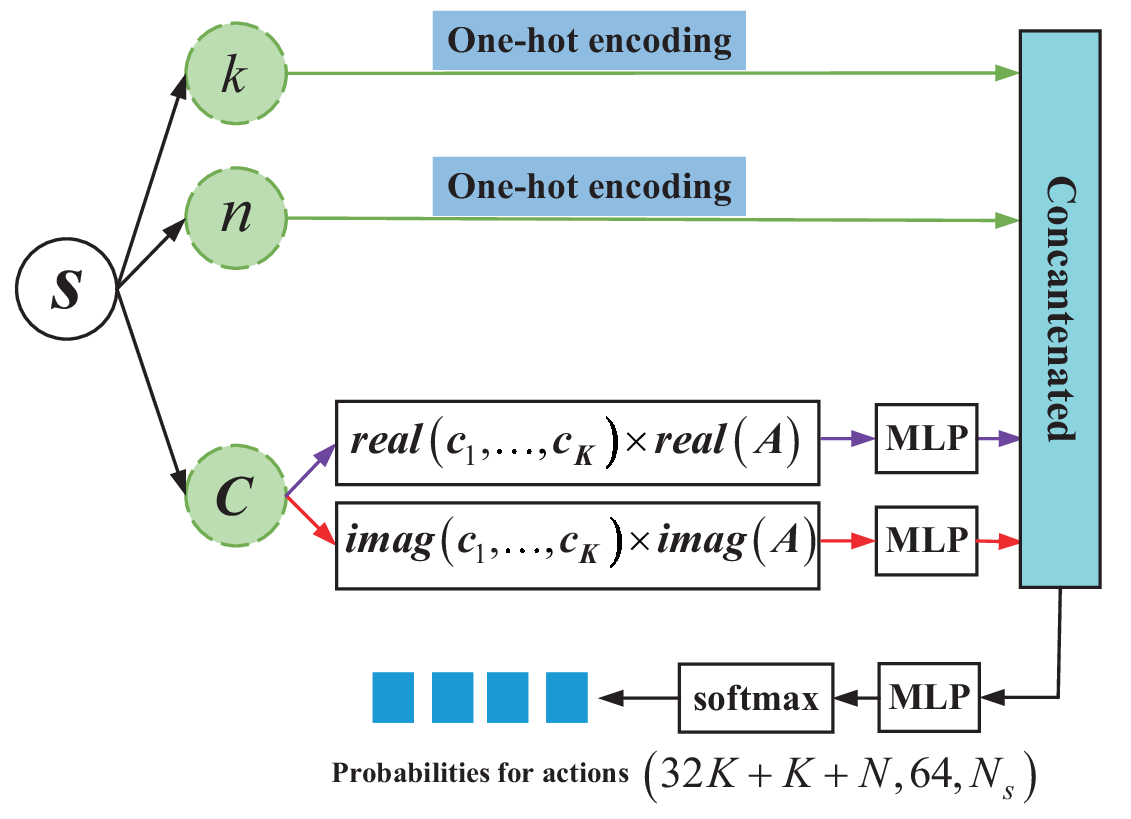}
	\caption{The structure of policy network.}
	\label{policymlp}
\end{figure}

Within the formulated MDP, the objective of the digital receiver is to obtain an optimal strategy that maximizes the aggregate reward by the time it reaches a terminal condition. The strategy, represented as $\bm{\pi}$, serves as a rule for selecting actions based on the current state, effectively creating a correspondence from the collection of all possible states $\mathcal{S}$ to the set of potential actions $\mathcal{A}$, expressed as $\bm{\pi}:\mathcal{S}\rightarrow \mathcal{A}$.
To establish what constitutes the best strategy $\bm{\pi}^*$, we first introduce the concept of the state value function $V^{\bm{\pi}}(\bm{s})$. This function, contingent upon a given policy $\bm{\pi}$ and parameter vector $\bm{w}$, quantifies the anticipated sum of rewards that expects to accumulate starting from a particular state s and continuing to act according to policy $\bm{\pi}$. Thus, the optimal policy $\bm{\pi}^*$ can be identified through this framework, aiming for the highest expected return from any initial state.
The state value function can be expressed as
\begin{equation}
	V(\bm{s}|\bm{\pi},\bm{w})= 
	\begin{cases}
		-L(\bm{C}_t,\bm{w})+\alpha \log_2(1+\text{SINR}), \\ \quad\quad\qquad\qquad \text{if $\bm{s}$ is terminal state}.\\
		V(\bm{s'}|\bm{\pi},\bm{w})|_{\bm{s'}=\mathcal{T}(\bm{s},\bm{\pi}(\bm{s}))}, \text{otherwise}.
	\end{cases}
\end{equation}


To solve problems (P1) and (P2), we adopted an algorithm that combines policy gradient based policy network with deep neural network-based sensing network. Among them, the policy network is responsible for exploring the optimal configuration of RIMSA, while the sensing network is complex and maps received signals to sensing results. The policy network and sensing network are jointly optimized based on the loss function until a satisfactory sensing effect can be achieved.


In the designed algorithm, the digital receiver commences from an initial condition denoted as $\bm{s}_0$ and follows a strategy that determines an action for every encountered state until arriving at a final state. For deciding on action within each state, the receiver employs a strategy $\bm{\pi}$ to convert the present state into a vector of probabilities. We utilize a neural network to represent the policy function, referred to as the policy network, with training accomplished through a policy gradient method. More precisely, the policy network is symbolized by $\bm{\pi}^{\bm{\theta}}(\bm{s}|\bm{w})$, wherein $\bm{\theta}$ signifies the set of parameters defining the policy network's configuration.


The structure of the policy network is depicted in Fig. \ref{policymlp}. 
For a given state $\bm{s}$, elements $k$ and $n$ are converted into $K$-dimensional and $N$-dimensional one-hot vectors, respectively. The determination of RF sensing outcomes within the target space is mainly influenced by the complex matrix $\bm{\vec{C}}$. Initially, the control matrix undergoes decomposition into its real and imaginary components, which are subsequently multiplied on the right by the corresponding real and imaginary parts of matrix $\bm{A}$. The outputs derived from these operations are then processed through a MLP.
Following this, the resultant one-hot vectors along with the $2K$ feature vectors extracted from each component $\bm{C}_1$ to $\bm{C}_K$ are concatenated. This combined data set is then input into the  MLP. 
For structured data, MLP can explore nonlinear features through their fully connected architecture when the sample size is limited. MLP is more resource-efficient, offering faster training and inference speeds, making them particularly suitable for edge devices or embedded systems. MLP maintains stable generalization performance through regularization techniques and relatively shallow architectures.
The output from this final MLP layer is channeled through a softmax function, yielding an $N_s$-dimensional vector. This vector represents the probabilities associated with selecting among $N_s$ possible actions.

\begin{figure}[!t]
	\centering
	\includegraphics[width=3.2in]{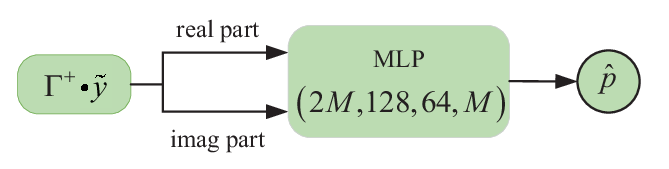}
	\caption{The structure of sensing network.}
	\label{sensemlp}
\end{figure}

\subsection{Sensing Outcomes Based on Sening Network}

We employ a neural network to realize the mapping from the received signal to the sensing outcome, as depicted in Fig. \ref{sensemlp}.
The received vector $\hat{\bm{y}}$ is left-multiplied by the pseudo-inverse of the matrix $\bm{\Gamma}_{Tx}$, denoted as $\bm{\Gamma}_{Tx}^+$ . The pseudo-inverse $\bm{\Gamma}_{Tx}^+$ can be computed using Singular Value Decomposition (SVD).
According to the least squares method, the vector $\bm{\hat{v}}=\bm{\Gamma}_{Tx}^+ \cdot\hat{\bm{y}}$ represents the optimal linear decoder for the actual reflection vector 
$\bm{\hat{v}}$ without jamming attack. Similarly, we also use this operation $\bm{\hat{v}}=\bm{\Gamma}_{Tx}^+ \cdot\hat{\bm{y}}_J$ to assist sensing, although this operation amplifies the impact of noise. This step potentially enhances the accuracy of the sensing network.

To proceed the vector $\bm{\hat{v}}$, we first decompose it into its real and imaginary parts. This decomposition is essential for handling complex-valued signals in a format that can be effectively processed by machine learning models, which typically operate on real numbers.
The real and imaginary parts are then stacked along the row direction to form a two-dimensional vector.
This newly formed vector serves as the input to the sensing network, which is composed of MLP networks.
The MLP consists of four fully connected layers.
The first three layers use the ReLU activation function and the final layer employs the softmax activation function, which converts the output of the network into a probability vector $\hat{\bm{p}}$.
This vector provides a clear representation of what the network senses about the objects whether empty or not.

\subsection{Training}
During training, it is essential to differentiate between a distributed RIMSA sensing scenario and a distributed RIMSA sensing scenario under jamming attack. In the absence of jamming attack, \eqref{P1} is used as the objective loss function, with the training dataset comprising only the received signals at the RIMSA units from the transmitter's directional signals. When jamming attack is present, \eqref{P2} serves as the objective loss function, and the training dataset includes the jamming signals emitted by the attacker in addition to the normal received signals.
Notably, in both scenarios, the policy network and the sensing network share the same network architecture for signal processing and network training. Correspondingly, for environments with and without jamming attack, the appropriate networks should be utilized to achieve effective sensing.

The training of the sensing network focuses on reducing the discrepancy between the predicted sensing results and the actual ground truth. This process ensures that the sensing network can reliably map received signals into expected sensing outcomes.
The sensing network undergoes training aimed at incrementally boosting the reward of the digital receiver. 
This method allows for continuous improvement of the sensing ability of the sensing network, thereby providing better reward values to support parameter updates of the policy network and make better decisions.

During the training phase of the policy network, a policy gradient approach is employed. A single training epoch is characterized by the sequence of state transitions that the RIMSA undergoes from the starting state until it reaches a terminal state. Throughout this epoch, the experiences, including states, actions, and rewards, are collected in a replay buffer for subsequent use in training. Once these experiences have been utilized for updating the network parameters, they are then removed from the buffer to make room for new data.
According to the policy gradient methodology, when training the network, the gradient of the state value function $V(\bm{s}_0|\bm{\pi},\bm{w})$ with respect to the parameters $\bm{\theta}$ satisfies
\begin{equation}\label{tidu}
	\nabla_{\bm{\theta}}V(\bm{s}_0 |\bm{\pi},\bm{w}) \propto \mathbb{E} \left[V(\mathcal{T}(\bm{S}_t,A_t)|\bm{\theta},\bm{w}) \frac{\nabla_{\bm{\theta}}\bm{\pi}_{A_t}^{\bm{\theta}}(\bm{S}_t|\bm{w}) }{\bm{\pi}_{A_t}^{\bm{\theta}}(\bm{S}_t|\bm{w})} \right].
\end{equation}

In equation \eqref{tidu}, $(\bm{S}_t,A_t)$ represents samples of states and actions from the replay buffer following policy $\bm{\pi}$. 
The state value function $V(\mathcal{T}(\bm{S}_t,A_t)|\bm{\pi},\bm{w})$ denotes the expected cumulative reward from selecting action $A_t$ in state $\bm{S}_t$.
To compute the gradient in equation \eqref{tidu}, it is necessary to calculate the reward for the digital receiver. If $s$ is a terminal state, the reward $R$ can be calculated through Monte Carlo methods, which in this context can be expressed as
\begin{equation}\label{reward}
	\begin{split}
		R(\bm{s}|\bm{w})= & -\sum_{\bm{v}\in\mathcal{V}}\sum_{i=1}^{N_{mc}}\bigg[\sum_{m=1}^{M}p_m(\bm{v})\cdot \ln(\hat{p}_m)+ \\
		& (1-p_m(\bm{v}))\cdot \ln(1-\hat{p}_m) \bigg]_{\bm{\hat{p}}=\bm{f}^{\bm{w}}(\hat{\bm{y}})}.\\
	\end{split}	
\end{equation}

If $s$ is a non-terminal state, the reward $R$ will be zero. 
In equation \eqref{reward}, $N_{mc}$ denotes the number of sampled noise vectors. 
Given that the reward is zero in non-terminal states, 
$V$ equals the reward at the final state under state $\bm{S}_t$, 
action $A_t$, and policy $\bm{\pi}^{\bm{\theta}}$.
In equation \eqref{reward}, the vector $\hat{\bm{p}}$ is generated by the sensing network.
The update of policy network is performed by the following 
\begin{equation}
	\bm{\theta}=\bm{\theta} + \alpha_{lr}\cdot  \mathbb{E} \left[V(\mathcal{T}(\bm{S}_t,A_t)|\bm{\theta},\bm{w}) \frac{\nabla_{\bm{\theta}}\bm{\pi}_{A_t}^{\bm{\theta}}(\bm{S}_t|\bm{w}) }{\bm{\pi}_{A_t}^{\bm{\theta}}(\bm{S}_t|\bm{w})} \right].
\end{equation}
where the gradient $\nabla_{\bm{\theta}}\bm{\pi}_{A_t}^{\bm{\theta}}(\bm{S}_t|\bm{w})$ is calculated by using the
back-propagation algorithm, and $\alpha_{lr}$ denotes the learning
rate.


For the training of the policy network, the goal is to update the parameters $\bm{\theta}$ to maximize the expected cumulative reward, while the sensing network is updated to minimize the loss between the network outputs and the true labels, leading to a coordinated optimization of both networks. This process helps to refine the policy and the sensing accuracy over time, ultimately achieving high-precision sensing even in the presence of the jamming attack.

\begin{table}
	\centering
	\caption{Simulation Parameters}  
	\begin{tabular}{|c|c|}	
		\hline
		\textbf{Parameter} & \textbf{Value} \\
		\hline
		Tx antenna gain $(g_T)$ &    $21.0\:$dBi \\
		\hline
		Tx power $(P)$ &    $100\:$mW \\
		\hline
		Number of RF chain $(N_{RF})$  &    $3\:$ \\
		\hline
		\makecell{Number of reconfigurable element \\ per RF chain $(N_{G})$}  &    $16\:$ \\
		\hline
		Signal frequency $f_c$ &    $3.198\:$GHz \\
		\hline
		\makecell{Number of each reconfigurable \\ element available states $(N_s)$} &   $4\:$ \\
		\hline
		Number of frames $(K)$ &    $20\:$ \\
		\hline
		\makecell{Probability of space grid \\ being nonempty $(p_{m,1})$} &    $0.5\:$ \\
		\hline
		Size of reflection vector set $(|\mathcal{V}|)$ &    $100\:$ \\
		\hline
		\makecell{Size of random sampled \\ subset $(Q^{sam}_{-m})$} &    $1000\:$ \\
		\hline
		Power of noise $(\epsilon)$ &    $10^{-9}\:$dBm \\
		\hline
		Size of space of interest $(l_x,l_y,l_z)$ &    $(0.1,0.1,0.1)\:$m \\
		\hline
		Initial learning rate $(\alpha_{lr})$ &    $0.001\:$ \\
		\hline
	\end{tabular}
	\label{para}
\end{table}


\section{Simulation Results}
In this section, we first describe the setup of the simulation scenario. The transmitter is located at $(0.87, -0.84, 0)$ m, and the receivers are placed on three walls at the center coordinates $(0, 2, 2)$ m, $(2, 2, 2)$ m, and $(1, 1, 3)$ m, with each RIMSA Rx considered as a single RF link. 
We set the $N_s$ discrete phase as $\{\pi/4,3\pi/4,5\pi/4,7\pi/4\}$.
The reflection coefficients of the reconfigurable elements under different configurations are simulated using CST, Microwave Studio, and the Transient Simulation Package \cite{reflectvvv} by assuming a vertically polarized RF signal with a 60-degree angle of incidence. The target space is a rectangular region, which is divided into $M$ spatial grid points, each with dimensions of $0.1 \times 0.1 \times 0.1$ $m^3$. Below the target space, at a distance, there is a jamming space set as a cube with a center coordinate of $(-1, 0, 0)$ m and dimensions of  $1 \times 1 \times 1$ $m^3$. During training, the jammer will randomly appear within this space. Upon completion of the training, the proposed sensing model will have the capability to resist interference within this space. The simulation parameters involved in the experiments are summarized in Tab. \ref{para}.

\subsection{Sensing Performance of Distributed RIMSA Rx}
\begin{figure}[!t]
	\centering
	\includegraphics[width=3.2in]{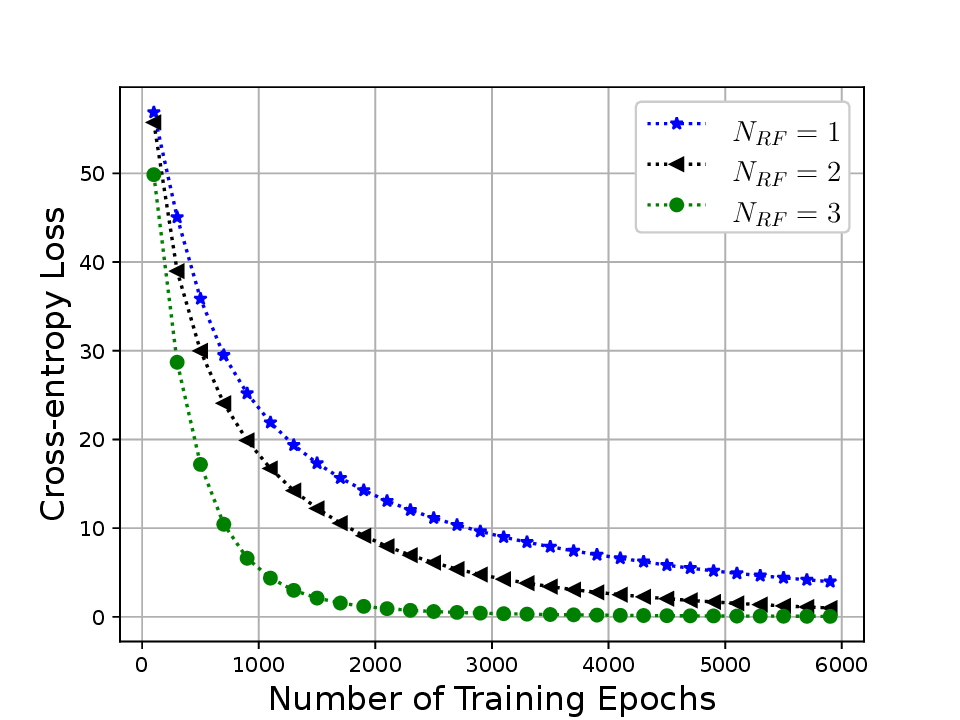}
	\caption{Cross-entropy loss versus the number of training epochs for different numbers of the RF chains.}
	\label{CEsense}
\end{figure}

\begin{figure}[!t]
	\centering
	\includegraphics[width=3.2in]{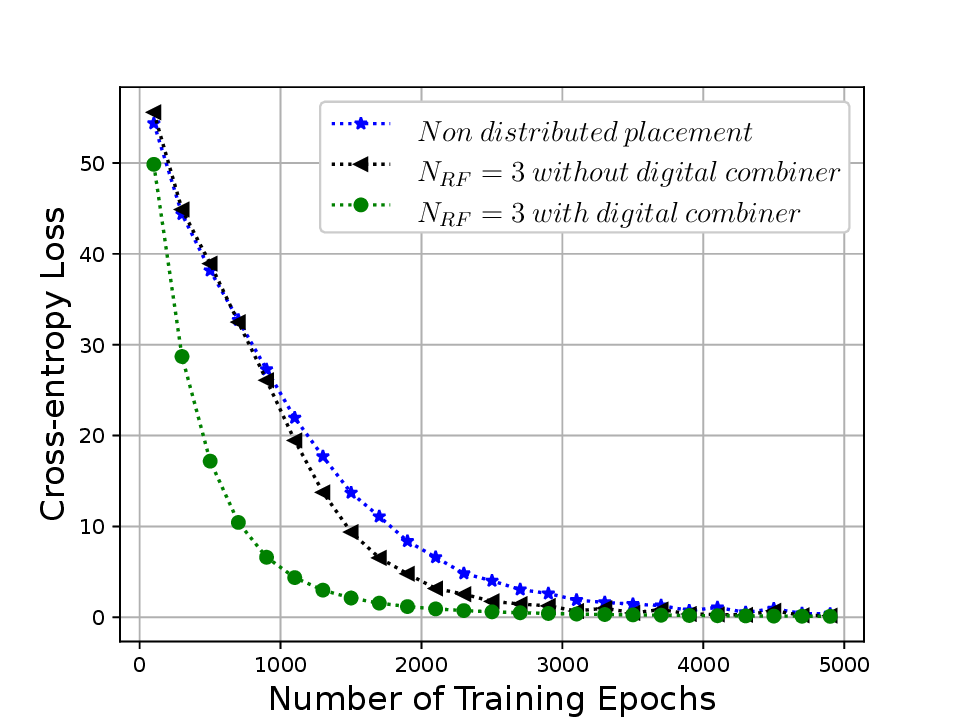}
	\caption{Cross-entropy loss versus the number of training epochs for different training methods.}
	\label{compare}
\end{figure}

\begin{figure}[!t]
	\centering
	\includegraphics[width=2.8 in]{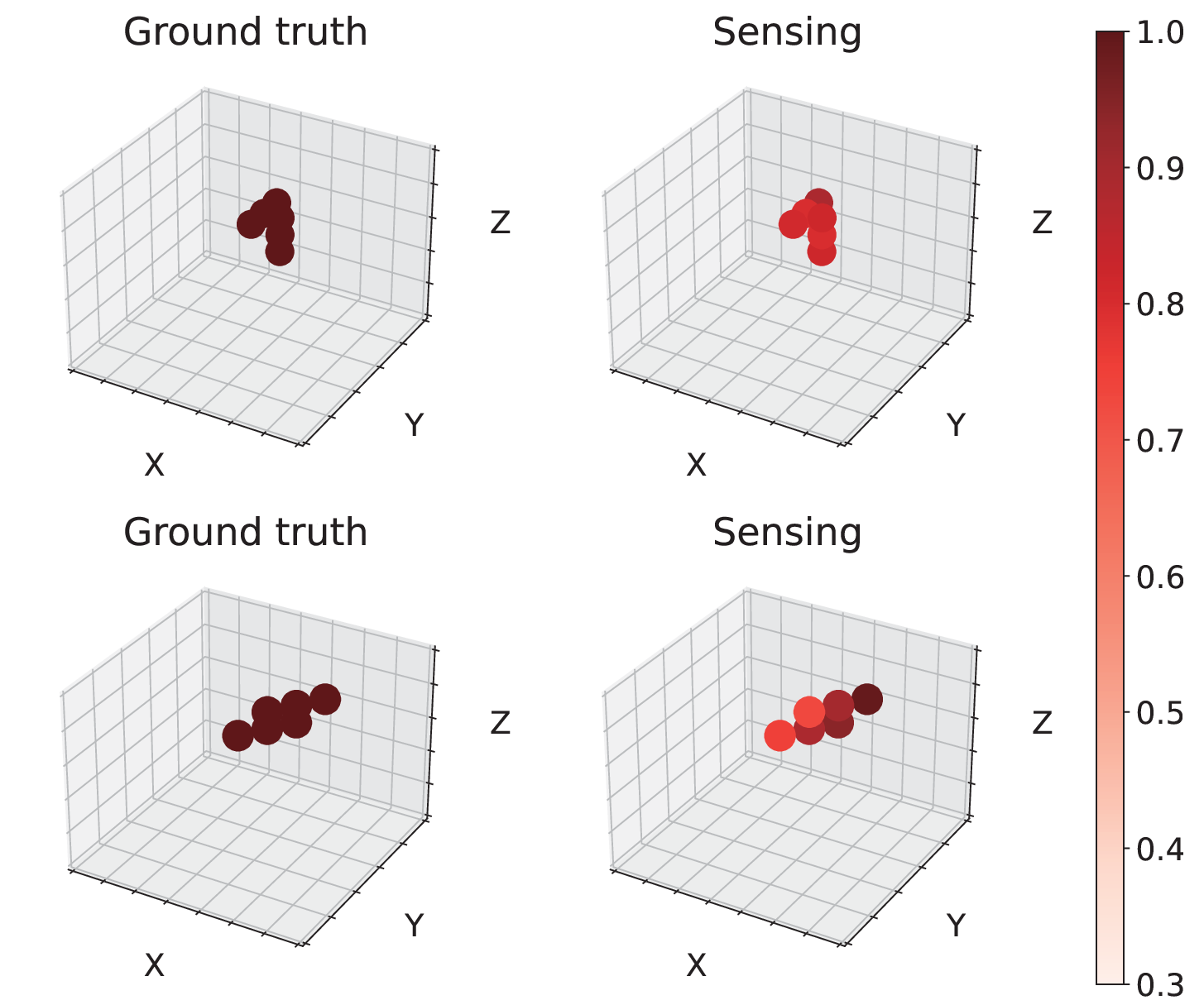}
	\caption{Comparison between ground-truth and the sensing results after training finished.}
	\label{3D_ground}
\end{figure}

\begin{figure}[!t]
	\centering
	\includegraphics[width=3.2in]{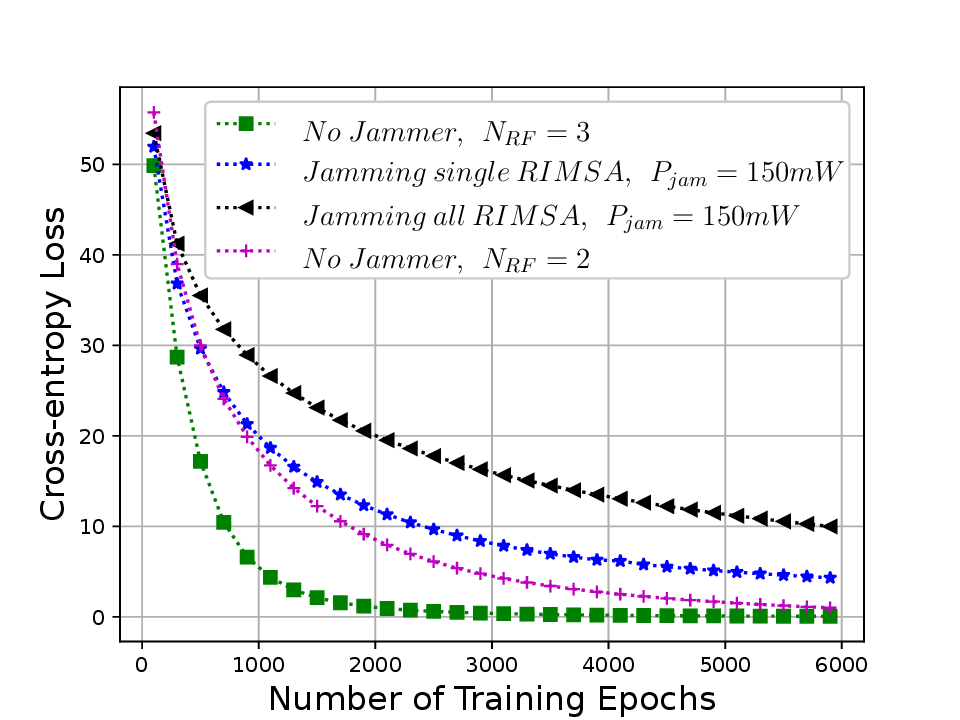}
	\caption{The impact of convergence performance on the training of sensing network under attack.}
	\label{RFANTIJAM}
\end{figure}

As illustrated in Fig. \ref{CEsense}, it becomes evident that an increase in the number of RF links facilitates faster convergence, leading to a minimal cross-entropy loss. 
With the increase of RF links, both strategic networks and sensing networks are capable of extracting a greater amount of information from received signals, thereby accelerating the convergence rate.
As illustrated in the Fig. \ref{compare}, the proposed distributed sensing approach exhibits markedly superior performance compared to both the method of placing all RIMSA elements together for sensing and the distributed placement of RIMSA elements without the use of a digital combiner. Benefiting from the distributed deployment and the gain provided by the digital combiner on the received signals, the proposed method demonstrates a significantly faster convergence rate, thereby showcasing its advantages over the other two approaches.

Fig. \ref{3D_ground} presents heatmaps comparing the ground truth with the sensing outcomes, where the color bar indicates the output probability of the sensing network. A higher probability signifies a greater likelihood of the presence of an object, represented by darker colors (note that the probabilities for cells without objects are not displayed in the figure). According to the experimental setup, if the output of the sensing network is greater than or equal to 0.5, it is determined that an object exists in the cell; conversely, a value less than 0.5 indicates the absence of an object. Upon completion of training, the sensing outcomes closely approximate the true values.

\begin{figure}[!t]
	\centering
	\includegraphics[width=3.2in]{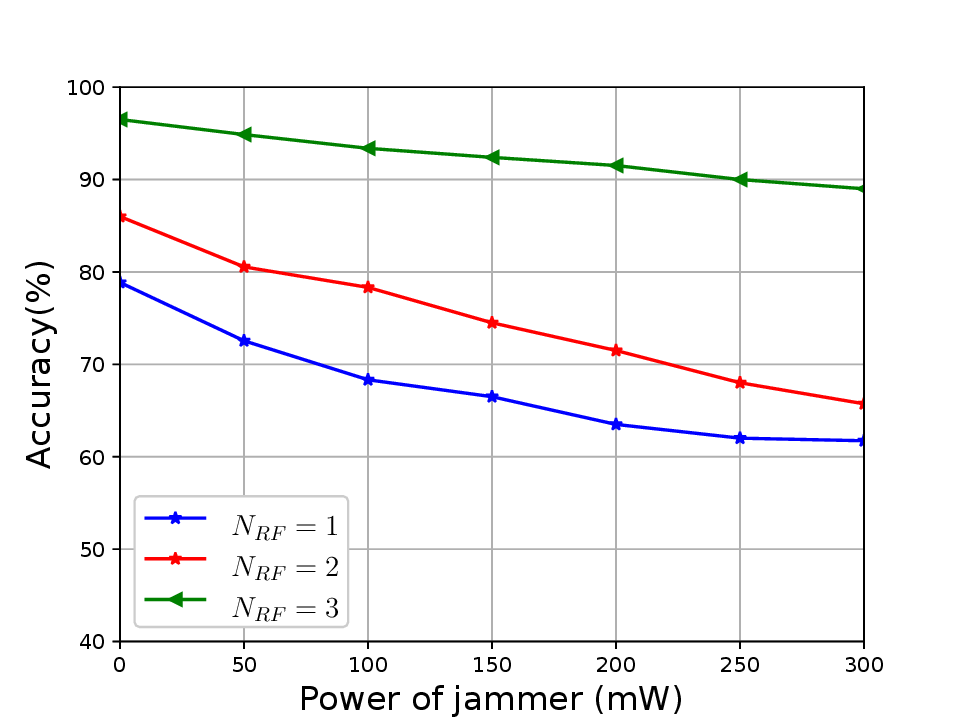}
	\caption{RF sensing accuracy versus the jamming powers with increasing RF chains.}
	\label{acc}
\end{figure}

\begin{figure}[!t]
	\centering
	\includegraphics[width=3.2in]{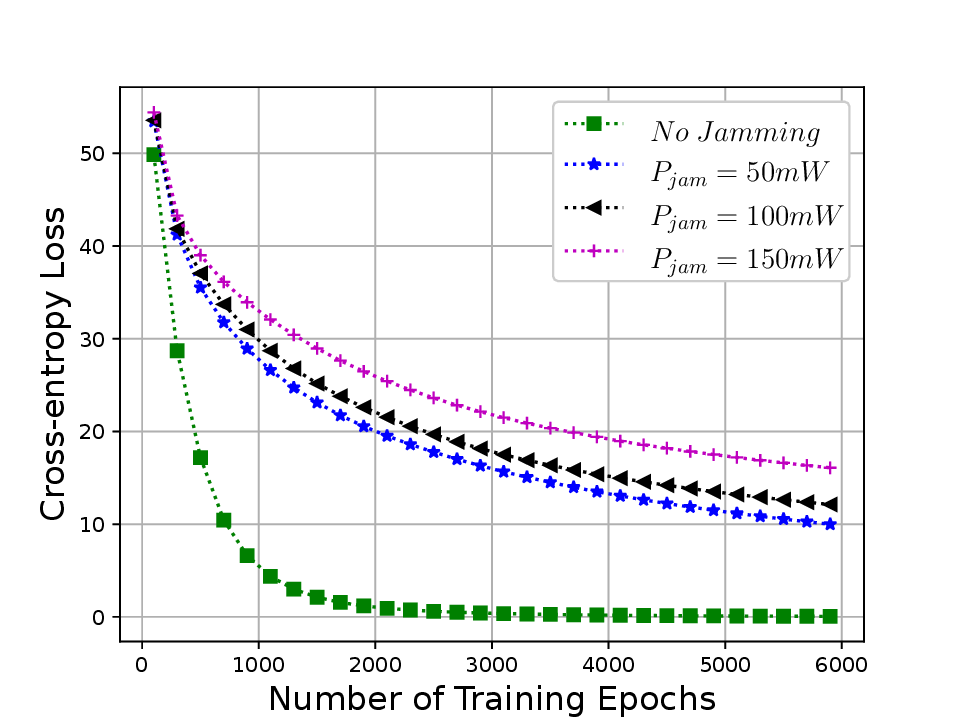}
	\caption{Cross-entropy loss versus the number of training epochs under different jamming powers with RF chains $N_{RF}=3$.}
	\label{Pjam}
\end{figure}

Fig. \ref{RFANTIJAM} illustrates the impact of different attack methods employed by the attacker on network training. When the attacker uses the same power to target a single RF chain, the detrimental effect is less severe compared to simultaneously attacking all links. This observation highlights the performance advantages afforded by distributed sensing schemes. In practical scenarios, where the attacker is often limited to links within close proximity, the distributed sensing architecture demonstrates a certain level of resilience against jamming attack.
Fig. \ref{acc} illustrates the sensing accuracy of the trained sensing model for real objects in test phase. Under the jamming attack, the sensing accuracy decreases as the jamming power increases, both for a single RF chain and for multiple RF chains. However, the sensing accuracy with three RF chains is significantly higher than that with a single RF link. This demonstrates the superiority of the distributed RIMSA-based sensing approach, which exhibits robustness against the jamming attack.
The enhanced performance and resilience of the system with multiple RF links highlight the potential of this approach for practical applications in complex and 
challenging wireless environments.

\begin{figure}[!t]
	\centering
	\includegraphics[width=3.2in]{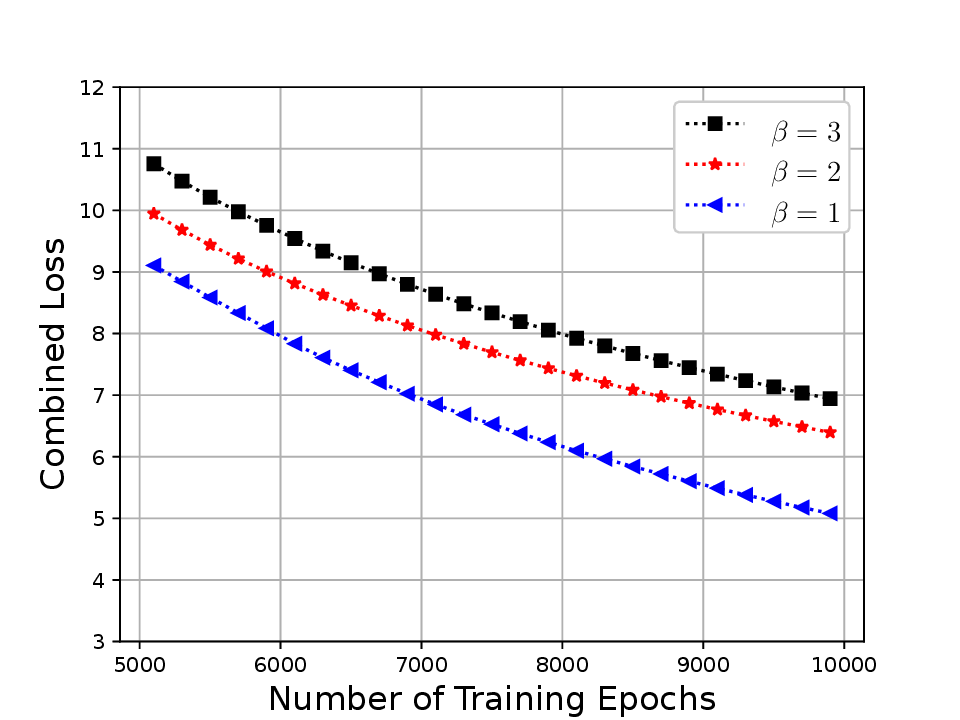}
	\caption{The impact of $\beta$ on anti-jamming RF sensing performance.}
	\label{beta}
\end{figure}

\begin{figure}[!t]
	\centering
	\includegraphics[width=3.2in]{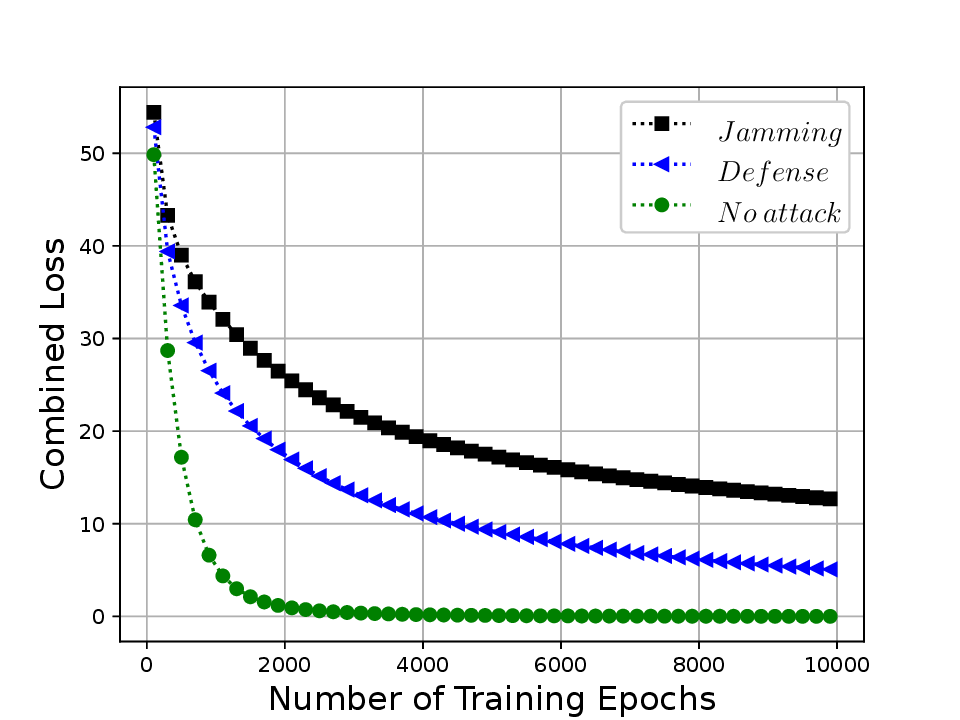}
	\caption{Combined loss versus the number of training epochs with defense  method at RF chains $N_{RF}=3$.}
	\label{defend}
\end{figure}

\subsection{Anti-Jamming Sensing Performance}

As illustrated in Fig. \ref{Pjam}, the performance of the original training method deteriorates sharply in the presence of jamming attack. As the jamming power increases, the training efficacy of the network worsens, with the loss function failing to decrease to the levels observed in the absence of attack, leading to a failure in achieving network convergence.

\begin{figure}[!t]
	\includegraphics[width=3.2in]{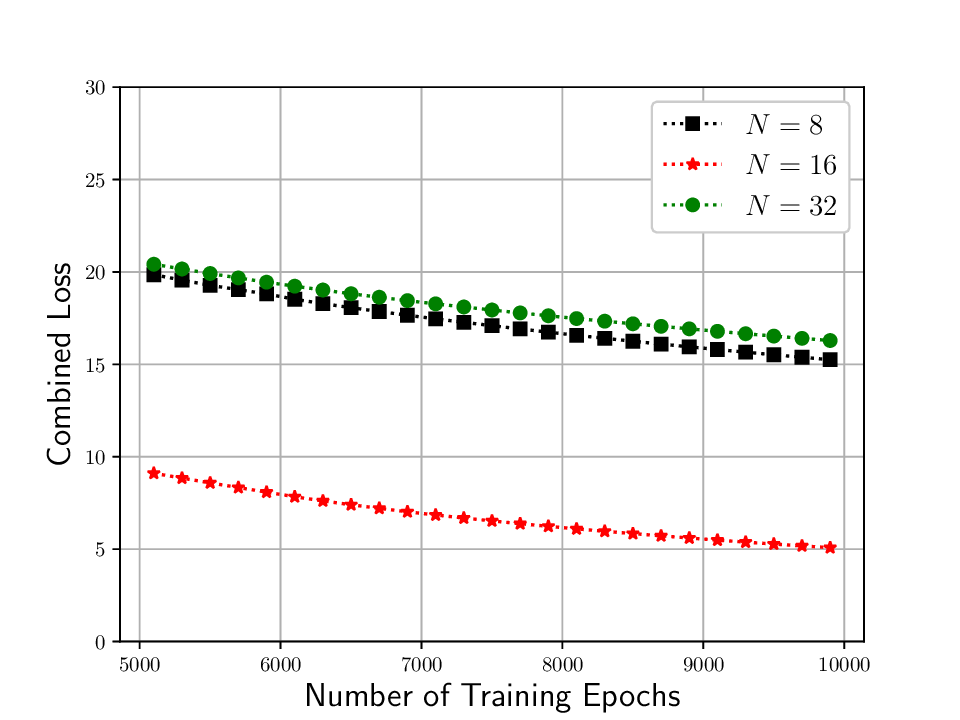}
	\caption{Anti-jamming performance vs the number of RIMSA elements.}
	\label{N}
\end{figure}

\begin{figure}[!t]
	\centering
	\includegraphics[width=3.2in]{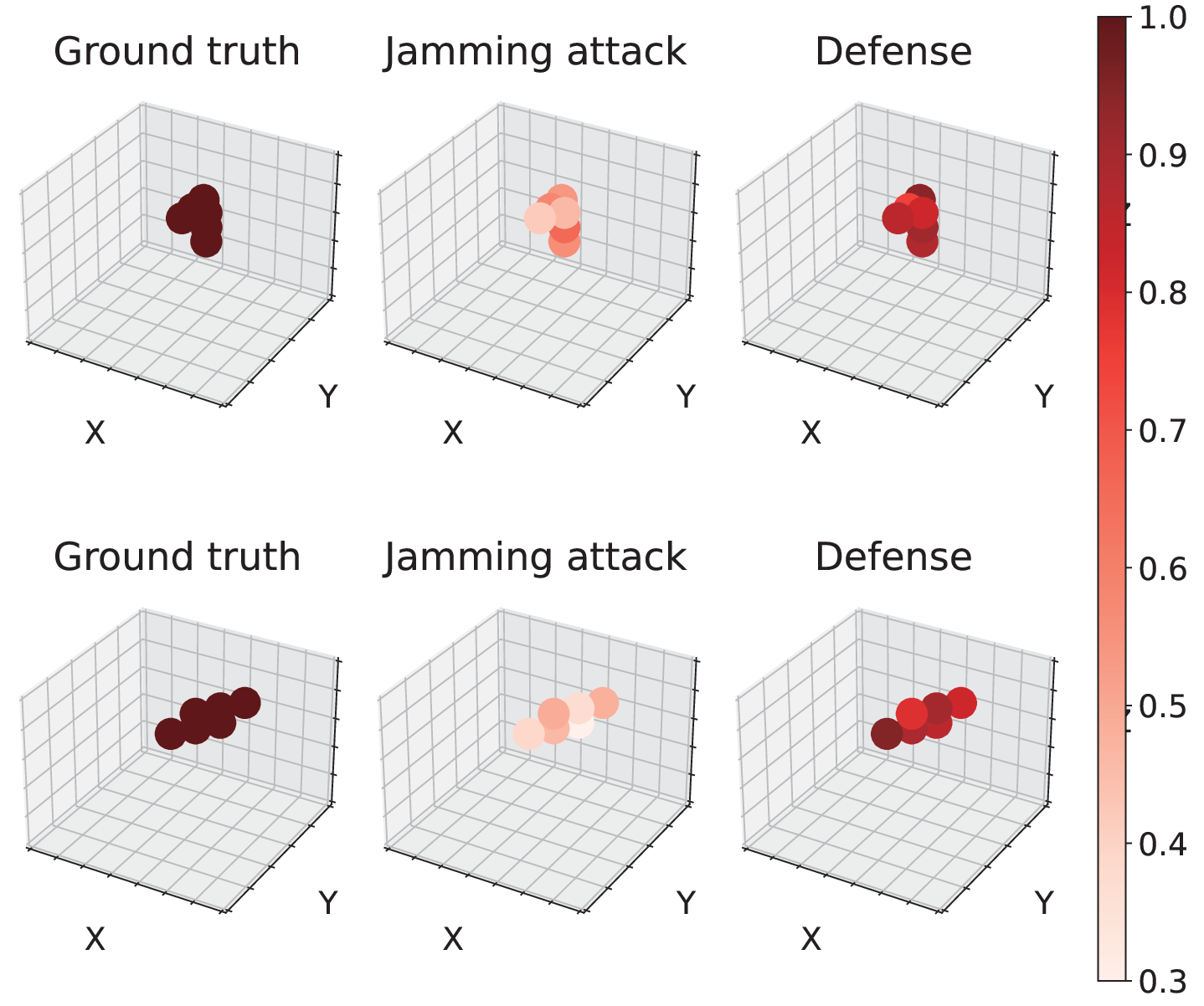}
	\caption{Illustrations of ground-truth and the sensing results after training finished with jamming attack exists.}
	\label{3D_jammer}
\end{figure}

To achieve high-precision sensing in environments with interference, we have redesigned the loss function used in network training. 
We introduce hyperparameter $\beta$ to balance the contributions of these loss components to RF sensing performance. As illustrated in Fig. \ref{beta}, a two-dimensional grid search methodically evaluates the impact of $\beta$ on network convergence. The final value is determined through a trade-off between computational complexity and sensing accuracy. In the subsequent experiments, we adopt $\beta=1$.
In Fig. \ref{defend}, it can be observed that the loss function plateaus under jamming attack and fails to decrease to the level observed in the absence of attack, leading to the non-convergence of the trained model and resulting in poor sensing performance. In contrast, the model trained with a loss function that incorporates the SINR is able to converge. 
That demonstrates the performance enhancement brought about by the newly designed loss function. Compared to the original cross-entropy loss function, the inclusion of the SINR factor effectively reduces the value of the loss, indicating that the new design is effective and facilitates network convergence.
When the number of elements per RF chain increases from $8$ to $16$, the anti-jamming performance improves and the network convergence accelerates, indicating enhanced anti-jamming capability with increased spatial degrees of freedom, as shown in Fig. \ref{N}. However, the growth of the state space and action space leads to degraded anti-jamming performance of the original network when the number continues to increase. The baseline network fails to effectively explore the enlarged state space, leading to reduced anti-jamming performance despite increased theoretical capability. This suggests there exists an limit for increasing elements of each RF chain. To pursue better anti-jamming performance, the network architecture must be modified accordingly.
Fig. \ref{3D_jammer} presents 3D sensing outcomes to illustrate the performance of interference-resistant sensing. The heatmap reveals that using the original network almost fails to achieve effective sensing, whereas the newly trained network is capable of accurately detecting the presence of objects.

\begin{figure}[!t]
	\centering
	\includegraphics[width=3.2in]{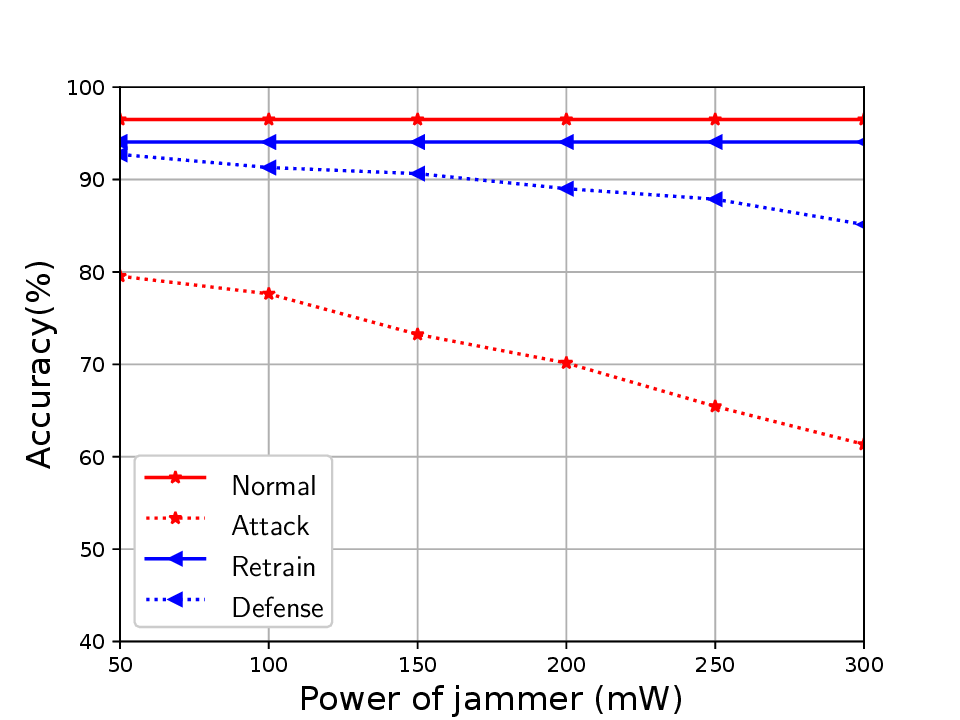}
	\caption{RF sensing accuracy versus the jamming powers under RF chains $N_{RF}=3$.}
	\label{acc_antijam}
\end{figure}

\begin{table}
	\centering
	\caption{The SINR(dB) performance of the received signals before and after anti-jamming methods.}  
	\begin{tabular}{|c|c|c|c|}	
		\hline
		Jamming power (mW) & $100$ & $200$ &$300$ \\
		\hline
		No defense  & $18.34$ & $15.45$ &$13.33$ \\
		\hline
		Zero-forcing  & $28.51$ & $24.37$ &$20.16$ \\	
		\hline
		Our scheme & $30.17$ & $27.49$ &$24.59$ \\	
		\hline
	\end{tabular}
	\label{sinr}
\end{table}

As shown in Fig. \ref{acc_antijam}, the accuracy of the retrained anti-jamming sensing model is slightly lower than that of the normally trained model in the absence of jamming attack. However,  the anti-jamming model trained with the combined loss function demonstrates a significantly higher accuracy under attack.
The sensing accuracy shows that as the jamming power increases, the original model is significantly affected, with its accuracy gradually decreasing. However, the model trained with the proposed method maintains a higher accuracy even as the jamming power increases.

Tab. \ref{sinr} reflects the anti-jamming performance of the received signals trained with a combined loss function from the perspective of SINR. 
We compared the performance of the proposed method with conventional multi-antenna systems employing zero-forcing algorithm for interference suppression. 
For fair comparison, both RIMSA assisted scenario and MIMO scenario adopt similar layouts as described in Section II. The single antenna transmitter emits directional probe signals, while the receiver is equipped with $N_{RF}$ antennas for signal reception. The receiver uses the zero-forcing algorithm to compute beamforming weights and determine the SINR of the received signals. The location of the attacker randomly appears in the designated area with different attacking powers. We calculate the average SINR of $1000$ received signals under each power.
Without any anti-jamming measures, the SINR of the received signals decreases as the jamming power increases. 
However, after undergoing training with a combined loss function that takes interference into account, the sensing network shows a marked improvement in its ability to resist interference. The SINR across different jamming power levels is improved and better than the traditional zero-forcing method. This signifies that the trained sensing network can maintain higher signal quality and achieve better performance even in the presence of jamming attack, effectively showcasing its enhanced robustness.

\section{Conclusion}

In this paper, we propose utilizing distributed RIMSA receivers for RF sensing. By programming their beamforming modes and employing digital combiner with MRC, we aim to overcome adverse environmental conditions and channel fading effects. The sensing problem is formulated as a joint optimization issue, involving the selection of beamforming pattern and the mapping of received signals to sensing outcomes. 
We jointly optimize the policy network with policy gradient algorithm to select beamforming pattern and the sensing network that maps received signals to sensing results. 
Furthermore, to enable sensing in the presence of jamming attack, we have designed a multi-factor loss function that takes into account the SINR of the received signals. Simulation results indicate that the distributed RIMSA can effectively perform sensing tasks, while the proposed combined loss function significantly enhances the precision of sensing under jamming attack.
In addition, we investigate RF sensing under a given number of RIMSA elements, where the network design is carefully matched to the corresponding state-action space. As the number of RIMSA elements increases, designing networks that can effectively exploit the anti-jamming capability remains an important research direction. Besides, we will focus on achieving robust RF sensing in the complex channel environments of 6G systems in future work.


\vfill

\end{document}